%% file: main.tex
\documentclass[preprint,12pt]{elsarticle}

\usepackage[T1]{fontenc}
\usepackage[utf8]{inputenc}
\usepackage{amsmath,amssymb,amsfonts,mathtools,bm}
\usepackage{graphicx}
\usepackage{booktabs}
\usepackage{array}
\usepackage{siunitx}
\usepackage{enumitem}
\usepackage{xcolor}
\usepackage{tikz}
\usetikzlibrary{arrows.meta,decorations.pathreplacing}
\usepackage[colorlinks=true,linkcolor=blue,citecolor=blue,urlcolor=blue]{hyperref}

\input{macros}

\journal{Computers and Fluids}

\begin{document}

\begin{frontmatter}

\title{Navier-Stokes characteristic boundary conditions for the Spectral
Difference method on simplex elements}

\author[cerfacs]{Thomas Marchal\corref{cor1}}
\ead{tmarchal@cerfacs.fr}
\cortext[cor1]{Corresponding author}
\author[cerfacs]{Maëlys Ruello}
\author[onera]{Hugues Deniau}
\author[cerfacs]{Guillaume Daviller}
\affiliation[cerfacs]{organization={CERFACS},
  addressline={42 avenue Gaspard Coriolis},
  postcode={31057}, city={Toulouse}, country={France}}
\affiliation[onera]{organization={ONERA/DMPE, Université de Toulouse},
  postcode={31055}, city={Toulouse}, country={France}}

\input{sections/00_abstract}

\end{frontmatter}

\input{sections/01_introduction}
\input{sections/02_governing}
\input{sections/03_sd}

\input{sections/04_nscbc_tensor}
\input{sections/05_nscbc_simplex}
\input{sections/06_results}
\input{sections/07_conclusions}

\section*{CRediT authorship contribution statement}
\textbf{Thomas Marchal:} Conceptualization, Investigation, Methodology,
Project administration, Software, Supervision, Validation,
Writing - original draft, Writing - review \& editing.
\textbf{Maëlys Ruello:} Conceptualization, Investigation, Methodology,
Software, Validation, Writing - original draft, Writing - review \& editing.
\textbf{Hugues Deniau:} Conceptualization, Investigation, Methodology,
Software, Supervision, Writing - review \& editing.
\textbf{Guillaume Daviller:} Investigation, Project administration, Software, Supervision, Validation,
Writing - review \& editing.

\section*{Declaration of competing interest}
The authors declare that they have no known competing financial interests
or personal relationships that could have appeared to influence the work
reported in this paper.

\section*{Data availability}
Data will be made available on request.

\section*{Funding}
This research did not receive any specific grant from funding agencies in
the public, commercial, or not-for-profit sectors.

\appendix
\input{sections/A_appendices}

\section*{Declaration of generative AI and AI-assisted technologies in the manuscript preparation process}
During the preparation of this work, the authors used Claude (Anthropic)
in order to improve the readability, clarity, and organization of the
text. After using this tool, the authors
reviewed and edited the content as needed and take full responsibility
for the content of the published article.

\bibliographystyle{elsarticle-num}
\bibliography{references}

\end{document}

%% file: macros.tex
\newcommand{\dd}[2]{\frac{\partial #1}{\partial #2}}
\newcommand{\T}{\mathsf{T}}                       %
\newcommand{\Jac}{|J|}                            %
\newcommand{\divx}{\nabla_{\bx}\cdot}             %
\newcommand{\divxi}{\nabla_{\bxi}\cdot}           %
\newcommand{\gradx}{\nabla_{\bx}}

\newcommand{\bx}{\mathbf{x}}
\newcommand{\bxi}{\bm{\xi}}
\newcommand{\bn}{\mathbf{n}}                      %
\newcommand{\bt}{\mathbf{t}}                      %
\newcommand{\hn}{\widehat{\mathbf{n}}}            %
\newcommand{\Mrot}{\mathsf{M}}                    %

\newcommand{\bU}{\mathbf{U}}
\newcommand{\bQ}{\mathbf{Q}}
\newcommand{\bW}{\mathbf{W}}
\newcommand{\bu}{\mathbf{u}}
\newcommand{\tU}{\widetilde{\mathbf{U}}}
\newcommand{\tQ}{\widetilde{\mathbf{Q}}}
\newcommand{\Ns}{N_s}                             %
\newcommand{\spec}{k}                             %
\newcommand{\tgt}{\mathrm{tgt}}                   %
\newcommand{\Lref}{L_{\mathrm{ref}}}              %
\newcommand{\Rgas}{r}                             %

\newcommand{\cF}{\bm{\mathcal{F}}}                %
\newcommand{\hcF}{\widehat{\bm{\mathcal{F}}}}     %
\newcommand{\bE}{\mathbf{E}}
\newcommand{\bF}{\mathbf{F}}
\newcommand{\bG}{\mathbf{G}}
\newcommand{\hE}{\widehat{\mathbf{E}}}
\newcommand{\hF}{\widehat{\mathbf{F}}}
\newcommand{\hG}{\widehat{\mathbf{G}}}
\newcommand{\En}{\mathbf{E}_{n}}                  %
\newcommand{\Et}[1]{\mathbf{E}_{t_{#1}}}          %
\newcommand{\tE}{\widetilde{\mathbf{E}}}          %
\newcommand{\tEs}{\widetilde{E}}                 %
\newcommand{\tF}{\widetilde{\mathbf{F}}}
\newcommand{\tG}{\widetilde{\mathbf{G}}}

\newcommand{\cL}{\bm{\mathcal{L}}}                %
\newcommand{\cT}{\bm{\mathcal{T}}}                %
\newcommand{\cN}{\bm{\mathcal{N}}}                %
\newcommand{\cS}{\bm{\mathcal{S}}}                %
\newcommand{\cA}{\bm{\mathcal{A}}}                %
\newcommand{\PU}{P_{\bU}}

\newcommand{\An}{A_{n}}
\newcommand{\RAn}{R_{A_n}}
\newcommand{\LAn}{L_{A_n}}
\newcommand{\LamAn}{\Lambda_{A_n}}

\newcommand{\NSP}{N_{\mathrm{SP}}}
\newcommand{\NFP}{N_{\mathrm{FP}}}
\newcommand{\NFPb}{N_{\mathrm{FP}}^{\partial}}    %
\newcommand{\Lsp}[1]{\ell^{\mathrm{SP}}_{#1}}     %
\newcommand{\Lfp}[1]{\ell^{\mathrm{FP}}_{#1}}     %
\newcommand{\bpsi}{\bm{\psi}}                     %
\newcommand{\RT}[1]{\mathbb{RT}_{#1}}
\newcommand{\Pp}[1]{\mathbb{P}_{#1}}
\newcommand{\DFP}{D^{\mathrm{FP}}}                %
\newcommand{\DSP}{D^{\mathrm{SP}}}                %
\newcommand{\Kb}{\mathcal{K}}                     %
\newcommand{\dK}{\partial\mathcal{K}_{\mathrm{B}}} %

%% file: sections/00_abstract.tex
\begin{abstract}
Unsteady compressible simulations with high-order methods on unstructured
meshes require boundaries that let acoustic, entropy and vorticity waves
leave the domain without spurious reflection. For the spectral difference
(SD) method, the Navier-Stokes characteristic boundary conditions (NSCBC)
have only been formulated on quadrilaterals and hexahedra. Their
strong-form formalism relies on a line of flux points normal to the
boundary face, which does not exist on triangles and tetrahedra. This paper extends the NSCBC to the SD method on simplex elements, including multi-species reacting gases.
The characteristic analysis is performed in the physical frame of each
flux point of the boundary face, and the normal derivative of the normal
flux is evaluated using the Raviart-Thomas divergence operator. The boundary fluxes are recovered from a linear system
local to each boundary element, whose matrix depends on the geometry only.  The treatment is compared with the tensor-product formulation at a comparable number of degrees of freedom on a series of test cases covering acoustic, vortical, viscous and reacting flows on triangles and tetrahedra. In all cases, the results are in close agreement with those obtained on quadrilaterals and hexahedra, so that the NSCBC become available for the SD method on simplex meshes, which are well suited to complex geometries.

\end{abstract}

\begin{keyword}
Spectral Difference \sep Raviart-Thomas \sep simplex elements \sep
Navier-Stokes characteristic boundary conditions \sep high-order methods \sep reacting flows
\end{keyword}

%% file: sections/01_introduction.tex
\section{Introduction}
\label{sec:intro}

High-order discontinuous methods combine low dissipation and dispersion
errors with a compact stencil on unstructured meshes
\cite{wang2013,vanharen2017}. The most widespread of them is the
discontinuous Galerkin (DG) method
\cite{cockburn2001,bassi1997,hesthaven2008}, which solves the weak form of
the governing equations. The equations are multiplied by test functions and
integrated over each element, and the resulting volume and surface integrals
are evaluated with quadrature rules whose cost increases with the
polynomial degree \cite{yu2014}. This cost has motivated the development of the DG spectral element method (DGSEM) \cite{kopriva2009}, where interpolation and quadrature nodes coincide and thus requires no additional
quadrature, at the price of aliasing errors that can destabilize nonlinear
computations \cite{gassner2013}. Split forms of the equations with the
summation-by-parts property make the DGSEM energy or entropy stable
\cite{gassner2013,carpenter2014,gassner2016}, and entropy-stable DGSEM
schemes are now used for compressible turbulent flows \cite{schwarz2025}.
The spectral difference (SD)
\cite{kopriva1996,kopriva1998,liu2006,may2006,sun2007} and flux
reconstruction (FR) \cite{huynh2007,wang2009,vincent2011} methods take
another route and solve the strong form, that is, the differential form of
the equations,
evaluated pointwise at the solution points (SP) within each element without any integration. In both
methods, the divergence is taken on a flux polynomial whose normal
component on the element faces equals the common interface flux: SD
interpolates the flux on a second set of points, the flux points (FP),
while FR adds correction functions to the discontinuous flux. For linear
equations, suitable correction functions make FR recover particular nodal
DG and SD schemes \cite{huynh2007,vincent2011}, but the schemes differ for
nonlinear equations \cite{cox2021}. Huynh extended the approach to
diffusion terms \cite{huynh2009}.

These methods are now used well beyond canonical test cases, for
turbulent, shocked, two-phase and reacting flows. The SD method has been
used for wall-modeled large-eddy simulation (LES) and subgrid-scale modeling
\cite{lodato2014,chapelier2016}, bypass transition
\cite{pinto2019,lodato2021}, aeroacoustics \cite{gao2017} and hypersonic
boundary-layer transition \cite{fievet2020hyp}. Shock-capturing techniques
\cite{lodato2019,tonicello2020,messai2024sc} and entropy-stable formulations
\cite{marchal2025} have extended it to shocked flows, such as the
supersonic Taylor-Green vortex \cite{chapelier2024} and supersonic jet
noise \cite{gao2026}. It has also been applied to turbomachinery
\cite{brunet2018}, two-phase flows \cite{tonicello2026}, detonations
\cite{gupta2018,tofaili2021} and premixed flames
\cite{marchal2023,castro2026}. The FR method has been applied to
turbomachinery \cite{lu2015,priebe2022,priebe2025} and to wall-modeled LES
on mixed meshes, up to high-lift configurations \cite{wang2022,wang2024}.
The DG method has been applied to hypersonic flows
\cite{ching2019,hoskin2024}, turbomachinery \cite{garai2016} and reacting
flows \cite{lv2014,billet2014,johnson2020}.

Industrial geometries are meshed automatically with simplex elements
(triangles and tetrahedra), so that a method intended for such
configurations must run on them. Energy-stable FR schemes have been derived on triangles
\cite{castonguay2012} and tetrahedra \cite{williams2014fr} for all orders of
accuracy. For SD, stability on simplex elements was obtained more recently,
by interpolating the flux in the Raviart-Thomas (RT) space
\cite{balan2012,veilleux2022jcp,veilleux2022jsc}, and the corrected scheme of
Messaï and Daviller \cite{messai2024} makes it stable for any order on
triangles and tetrahedra.

Unsteady compressible simulations also require boundaries that let the
acoustic, entropy and vorticity waves leave the domain without spurious
reflection. The Navier-Stokes characteristic boundary conditions (NSCBC) of
Poinsot and Lele \cite{poinsot1992}, built on the characteristic analysis of
Thompson \cite{thompson1987}, are the standard approach. They relax the
incoming waves towards target values with a linear relaxation term
\cite{rudy1980}, whose coefficient sets the acoustic behavior of the
boundary \cite{selle2004,polifke2006}. They have been extended to
multi-species reacting flows \cite{baum1995,sutherland2003}, to
multidimensional flows through transverse terms
\cite{yoo2005,yoo2007,lodato2008}, to generalized coordinates
\cite{kim2000,kim2004} and to unstructured grids \cite{granet2010}, and
inlets have been adapted to inject vortical and acoustic waves
\cite{daviller2019}. Fosso et al. \cite{fosso2012} compare several of these
outflow conditions for aeroacoustic simulations.

In high-order discontinuous methods, the boundary condition must act on the
discrete operator of the scheme. Fiévet et al. \cite{fievet2020} proposed a
compact formalism in strong form for the SD and FR methods on
tensor-product elements (quadrilaterals and hexahedra), where the correction of the boundary-normal flux
derivative is carried by the line of FP normal to the face, and
Marchal et al. \cite{marchal2023} extended it to multi-species reacting
flows. In DG methods, which solve the weak form, the characteristic
conditions enter instead through the external state of the boundary
numerical flux: with mirror
elements in DG \cite{toulopoulos2011}, with a polynomial correction of the
boundary state in DG on quadrilaterals \cite{shehadi2024}, and recently in
hybridizable DG on simplex elements \cite{ellmenreich2026}.

The strong-form formalism of Fiévet et al. does not carry over to simplex
elements. It relies
on a reference direction normal to the face, along which the flux derivative
is computed separately from the rest of the divergence, and on a line of
FP normal to the face, along which the correction is
one-dimensional. On triangles and tetrahedra, neither exists: the RT
operator only returns the full divergence of the flux, and no line of FP
runs normal to a face (see Section~\ref{sec:sd}). Fiévet et al. leave
tetrahedral elements to future work \cite[Sec.~4.1.2]{fievet2020}.
To the authors' knowledge, no NSCBC treatment has been proposed for the SD
method, nor for the FR method, on simplex elements.

In this work, the NSCBC are extended to the SD method on triangles and
tetrahedra. The characteristic analysis is performed in a frame attached to
the boundary face, the normal derivative of the normal flux is evaluated
with the RT divergence operator, and the boundary fluxes are recovered
through a small linear system local to each boundary element, which only
depends on the geometry. The treatment applies to single-species gases and to multi-species reacting mixtures, for which the face-frame analysis gains one wave per species, advected at the normal velocity (see~\ref{app:ms-nscbc}). This treatment is assessed on two- and
three-dimensional cases of increasing complexity, from an isentropic vortex
leaving through a straight or an inclined outlet to a Poiseuille flow in a
channel, a spherical pressure wave in a cube, a Poiseuille flow in a pipe
and a premixed methane-air flame in a multi-species reacting gas. Each case is compared with the tensor-product formulation of
Fiévet et al. at a comparable number of degrees of freedom (DoF).

The paper is organized as follows. Section~\ref{sec:governing} gives the
governing equations, and Section~\ref{sec:sd} recalls the SD method on
tensor-product and simplex elements. Section~\ref{sec:nscbc-tensor} recalls
the NSCBC formalism of Fiévet et al. on tensor-product elements and points
out the two properties on which it relies. Section~\ref{sec:nscbc-simplex}
presents the treatment on simplex elements, and Section~\ref{sec:results}
assesses it on the test cases above. Section~\ref{sec:conclusions} draws the
conclusions, and \ref{app:multispecies} gives the extension to multi-species
reacting flows.

%% file: sections/02_governing.tex
\section{Governing equations}
\label{sec:governing}

The three-dimensional compressible Navier-Stokes equations are solved for
a single-species, non-reacting gas, and read in the physical space
$\bx=(x,y,z)^\T$
\begin{equation}
  \dd{\bU}{t} + \divx\cF = \mathbf{0},
  \qquad
  \bU = \big(\rho,\ \rho u,\ \rho v,\ \rho w,\ \rho E\big)^\T,
  \label{eq:nse-phys}
\end{equation}
where $\rho$ is the density, $\bu=(u,v,w)^\T$ the velocity and $E$ the total
energy per unit mass. The flux tensor $\cF=(\bE,\bF,\bG)^\T$ gathers the
fluxes along $x$, $y$ and $z$, so that
$\divx\cF=\partial_x\bE+\partial_y\bF+\partial_z\bG$, and is split into a
convective and a diffusive part, $\cF=\cF_c+\cF_d$, whose rows read
\begin{equation}
  \cF_c =
  \begin{pmatrix}
    \rho\,\bu^\T\\
    \rho\,\bu\,\bu^\T + P\,I\\
    (\rho E+P)\,\bu^\T
  \end{pmatrix},
  \qquad
  \cF_d =
  \begin{pmatrix}
    \mathbf{0}^\T\\
    -\bm{\tau}\\
    \mathbf{q}^\T-(\bm{\tau}\bu)^\T
  \end{pmatrix},
  \label{eq:fluxes}
\end{equation}
with $P$ the static pressure and $I$ the identity matrix. The viscous
stress tensor and the heat flux are
\begin{equation}
  \bm{\tau} = \mu\big(\gradx\bu+\gradx\bu^\T\big)-\frac{2}{3}\,\mu\,(\divx\bu)\,I,
  \qquad
  \mathbf{q} = -\lambda\,\gradx T,
  \label{eq:tau-q}
\end{equation}
where $T$ is the temperature, $\mu$ the dynamic viscosity and $\lambda$ the
thermal conductivity. The gas is calorically perfect,
\begin{equation}
  P = \rho\,\Rgas\,T,
  \qquad
  \rho E = \frac{P}{\gamma-1}+\frac{1}{2}\rho\lVert\bu\rVert^2,
  \qquad
  c = \sqrt{\gamma P/\rho},
  \label{eq:closure}
\end{equation}
with $\Rgas$ the specific gas constant, $\gamma$ the heat capacity ratio and
$c$ the speed of sound. The thermal conductivity is
$\lambda=\mu C_p/\mathrm{Pr}$, with $C_p=\gamma\Rgas/(\gamma-1)$ the heat
capacity at constant pressure and $\mathrm{Pr}$ a constant Prandtl number.
The dynamic viscosity is constant in the viscous single-species cases of
Section~\ref{sec:results}, where its value is given.

The NSCBC theory uses the vector of primitive variables, which is
introduced here as $\bQ=(\rho,u,v,w,P)^\T$. The extension to a
multi-species reacting mixture, used in the burner case of
Section~\ref{sec:res-burner}, is given in \ref{app:multispecies}.

%% file: sections/03_sd.tex
\section{Spectral difference method}
\label{sec:sd}

This section recalls the SD discretization on tensor-product and simplex
elements, and isolates the discrete operator on which an NSCBC acts. This
operator is a directional flux derivative on tensor-product elements and
the RT divergence on simplex elements, and the difference
between the two drives the rest of the paper.

\subsection{Isoparametric transformation}
\label{sec:sd-iso}

Each physical element is the image, through a polynomial mapping, of a
reference element $\Kb$, which is the unit square or cube for
tensor-product elements and the unit triangle or tetrahedron for simplex
elements. The mapping and its Jacobian matrix read
\begin{equation}
  \bx(\bxi) = \sum_{i=1}^{N_g} M_i(\bxi)\,\bx_i,
  \qquad
  J = \dd{\bx}{\bxi},
  \label{eq:iso-map}
\end{equation}
where $\bx_i$ are the $N_g$ geometric nodes of the element, $M_i$ the shape
functions and $\bxi=(\xi,\eta,\zeta)^\T$ the reference coordinates, and $\Jac$
denotes the determinant of $J$.

The equations are solved in the reference element, where the chain rule
and the geometric conservation law \cite{kopriva2006,abe2015} turn
Eq.~\eqref{eq:nse-phys} into
\begin{equation}
  \Jac\,\dd{\bU}{t} + \divxi\hcF = \mathbf{0},
  \qquad
  \hcF = \Jac\,J^{-1}\cF \equiv \big(\hE,\hF,\hG\big)^\T,
  \label{eq:nse-ref}
\end{equation}
where $\divxi\hcF = \partial_\xi\hE+\partial_\eta\hF+\partial_\zeta\hG$, the
time derivative being kept multiplied by $\Jac$ as in~\cite{veilleux2022jcp} \footnote{Note that Fiévet et al.~\cite{fievet2020} use the
inverse definition of $\Jac$, so that their equations are multiplied by
$1/\Jac$.}.

The same identity provides the physical gradient of any scalar field $\phi$
from a reference divergence,
\begin{equation}
  \gradx\phi = \frac{1}{\Jac}\,\divxi\!\Big(\Jac\,J^{-\T}\phi\Big),
  \label{eq:grad-from-div}
\end{equation}
where the divergence acts on each row of the matrix $\Jac J^{-\T}\phi$.
This identity is used in Section~\ref{sec:simplex-dn} to evaluate the
boundary-normal derivative of the flux on simplex elements.

\subsection{On tensor-product elements (1D example)}
\label{sec:sd-tensor}

The scheme is presented on a one-dimensional element of degree $p$, which
yields an order of accuracy $p+1$. The element carries $p+1$ SP, where the conservative variables are stored and advanced in
time, located at the Gauss-Chebyshev points of the first kind $\xi_j$,
$j\in[1,p+1]$ \cite{marchal2023}. It also carries $p+2$ FP, namely the $p$
Gauss-Legendre points and the two end points $\xi=0$ and $\xi=1$. The
stability of the SD scheme does not depend on the location of the SP
\cite{vandenabeele2008}, and placing the interior FP at the Gauss-Legendre
points makes it stable for all orders of accuracy on tensor-product
elements \cite{jameson2010}.

The solution is interpolated at the FP, where the fluxes are evaluated.
With $\Lsp{l}$ and $\Lfp{l}$ the Lagrange polynomials based on the SP and on the
FP, the solution and flux polynomials read
\begin{align}
  \overline{\bU}(\xi) &= \sum_{l=1}^{p+1}\bU_l\,\Lsp{l}(\xi),
  \label{eq:sd1d-U}\\
  \overline{\hE}(\xi) &= \sum_{l=1}^{p+2}\hE_l\,\Lfp{l}(\xi).
  \label{eq:sd1d-E}
\end{align}
The flux polynomial is then differentiated at the SP,
\begin{equation}
  \left(\dd{\hE}{\xi}\right)_{\!j} = \sum_{l=1}^{p+2}\hE_l\,\dd{\Lfp{l}}{\xi}(\xi_j),
  \qquad j\in[1,p+1],
  \label{eq:sd1d-dE}
\end{equation}
which closes the semi-discrete scheme. At the two end points, the flux
entering Eq.~\eqref{eq:sd1d-dE} is replaced by a common interface value, the numerical flux, which couples the element to its neighbors. Its convective part is given by a Riemann solver. The diffusive part, on the other hand, is computed  following the approach of Bassi and Rebay~\cite{bassi1997}, applied to SD by Sun et al.~\cite{sun2007}: at each interface FP, the solution and its gradient are averaged between the two simplex elements\footnote{The diffusive part is a numerical viscous flux not to be confused with the physical diffusive flux $\cF_d$ of Eq.~\eqref{eq:fluxes}}.

\begin{figure}[htbp]
  \centering
  \includegraphics[width=0.6\linewidth]{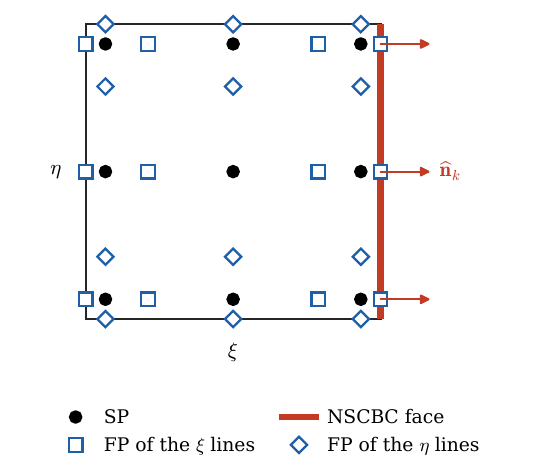}
\caption{SP and FP in the reference quadrilateral for $p=2$. The FP lie
on lines along $\xi$ and $\eta$, and each FP of the NSCBC face closes a
line normal to it.}
\label{fig:sp-fp-quad}
\end{figure}

In two and three dimensions, the SP and FP follow from a tensor product. Figure~\ref{fig:sp-fp-quad} shows them on a quadrilateral for $p=2$, where the FP lie on lines along $\xi$ and $\eta$. Each line of FP along $\xi$ carries $\hE$ and yields $\partial_\xi\hE$ through Eq.~\eqref{eq:sd1d-dE}, and likewise for
$\eta$ and $\zeta$, so that the three terms of $\divxi\hcF$ are computed
separately. Moreover, at a face $\xi=1$, each boundary FP closes a line of FP
normal to that face. Section~\ref{sec:nscbc-tensor} relies on both
properties.

\subsection{On simplex elements}
\label{sec:sd-simplex}

On simplex elements, the original SD formulation \cite{liu2006} is weakly
unstable at third and fourth order \cite{vandenabeele2008}. The
tensor-product construction is therefore replaced by the
spectral difference Raviart-Thomas (SDRT) scheme
\cite{veilleux2022jcp,veilleux2022jsc}, which builds on the RT flux
interpolation of Balan et al. \cite{balan2012}. The
solution $\bU_h$ is a polynomial of degree $p$ defined at the $N_{SP}$ solution points of Eq.(~\ref{eq:rt-nsp}):
\begin{equation}
  \NSP = \frac{(p+1)(p+2)}{2}\ \text{(2D)},
  \qquad
  \NSP = \frac{(p+1)(p+2)(p+3)}{6}\ \text{(3D)}
  \label{eq:rt-nsp}
\end{equation}
SP are located at the points of symmetric quadrature rules: those of Williams, Shunn and Jameson on triangles~\cite{williams2014}, and those of Shunn and Ham on tetrahedra~\cite{shunn2012}. Since no closed-form Lagrange basis exists for an arbitrary set of points on a simplex, the solution is expanded in the orthogonal modal basis of Proriol, Koornwinder
and Dubiner~\cite{koornwinder1975,dubiner1991}, and interpolated at the FP through a generalized Vandermonde matrix~\cite{veilleux2022jcp}.

\begin{figure}[htbp]
  \centering
  \includegraphics[width=0.6\linewidth]{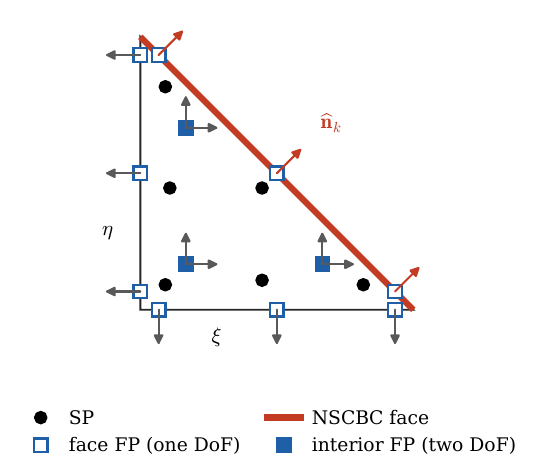}
  \caption{SP and FP in the reference triangle with the SDRT scheme for
  $p=2$. Face FP carry the face normal $\widehat{\mathbf{n}}_k$, interior
  FP carry two DoF, and no line of FP is normal to the NSCBC face.}
  \label{fig:sp-fp-tri}
\end{figure}

The flux is approximated as a single vector field in the RT
space $\RT{p}$, the smallest polynomial space whose divergence lies in
$\Pp{p}$. Its expansion reads
\begin{equation}
  \hcF_h(\bxi) = \sum_{k=1}^{\NFP}\big(\hcF_k\cdot\hn_k\big)\,\bpsi_k(\bxi),
  \label{eq:rt-flux}
\end{equation}
with a number of flux DoF:
\begin{equation}
  \NFP = (p+1)(p+3)\ \text{(2D)},
  \qquad
  \NFP = \frac{(p+1)(p+2)(p+4)}{2}\ \text{(3D)}.
  \label{eq:rt-nfp}
\end{equation}
The basis functions $\bpsi_k$ satisfy the interpolation property
\begin{equation}
  \bpsi_j(\bxi_k)\cdot\hn_k = \delta_{jk},
  \label{eq:rt-prop}
\end{equation}
where $\delta_{jk}$ is the Kronecker symbol and $\hn_k$ the vector
attached to the DoF $k$. On a face, $\hn_k$ is normal to that
face in $\Kb$ and points outwards. Its length is a convention, for instance
$\hn_k=(1,1)^\T$ on the hypotenuse of the reference triangle in our
implementation, with which Eq.~\eqref{eq:rt-prop} makes the scheme
consistent. Inside the element, each physical FP carries $d$ DoF, one per
reference unit vector, with $d$ the space dimension.
Figure~\ref{fig:sp-fp-tri} shows these points and vectors on a triangle for
$p=2$. The face FP are located at the Gauss-Chebyshev
points of the first kind on triangle edges, and at the SP of a triangle of the same degree on tetrahedron faces. 
Linear stability analysis is used to define the location of the interior FP: the single interior point is the centroid for $p=1$, and the interior points coincide with the SP of degree $p-1$ for $p=2$ on tetrahedra~\cite{veilleux2022jsc} and for $p=2$ and $p=3$ on triangles~\cite{veilleux2022jcp}. Stable sets of interior FP were obtained by optimization for $p=4$ and $p=5$ on triangles~\cite{veilleux2022jcp}, whereas none is known for $p>2$ on tetrahedra~\cite{veilleux2022jsc} with the SDRT method.

The scalar coefficients of Eq.~\eqref{eq:rt-flux} come from the physical
flux, as its projection inside the element and as the numerical flux of
Section~\ref{sec:sd-tensor} on a face, which gathers the convective and the
diffusive contributions:
\begin{equation}
  \hcF_k\cdot\hn_k =
  \begin{cases}
    \cF\big(\bU_h(\bxi_k)\big)\cdot\Jac_k J_k^{-\T}\hn_k, & \bxi_k\in\Kb\setminus\partial\Kb,\\[2pt]
    \big(\cF\cdot\Jac_k J_k^{-\T}\hn_k\big)^{\mathrm{num}}, & \bxi_k\in\partial\Kb .
  \end{cases}
  \label{eq:rt-coef}
\end{equation}
The divergence of Eq.~\eqref{eq:rt-flux} at the SP closes the
semi-discrete scheme. For $j\in[1,\NSP]$,
\begin{equation}
  \Jac_j\,\dd{\bU_j}{t} + \sum_{k=1}^{\NFP}\DSP_{jk}\,\big(\hcF_k\cdot\hn_k\big) = \mathbf{0},
  \qquad
  \DSP_{jk} = \big(\divxi\bpsi_k\big)(\bxi_j).
  \label{eq:rt-div}
\end{equation}

Equation~\eqref{eq:rt-div} returns the full divergence $\divxi\hcF$ at each
SP and, unlike Eq.~\eqref{eq:sd1d-dE}, does not split into directional
derivatives. In addition, no line of FP runs normal to a face
(Fig.~\ref{fig:sp-fp-tri}), so that the two properties on which the
tensor-product NSCBC treatment relies are both lost on simplex elements.

The stability of the SDRT scheme depends on the location of the interior FP,
and no stable set is known for every $p$ \cite{veilleux2022jcp,messai2024}.
Messaï and Daviller \cite{messai2024} remove this restriction with the
corrected scheme C-SDRT, which corrects the flux at the interior FP to compensate the error introduced by the numerical fluxes on the faces, and is stable for any set of FP on triangles and tetrahedra and for all $p$. The NSCBC treatment below is written for SDRT and applies unchanged to C-SDRT.

%% file: sections/04_nscbc_tensor.tex
\section{NSCBC on tensor-product elements}
\label{sec:nscbc-tensor}

The NSCBC approach~\cite{thompson1987,poinsot1992} corrects the
boundary-normal flux derivative so that the waves entering the domain
satisfy the physical boundary conditions, while the waves leaving the
domain keep the values computed from the interior. On a tensor-product element, the
boundary is a face of constant reference coordinate, say $\xi=1$, and the
correction acts on $\partial_\xi\hE$.

At such a face, Fiévet et al.~\cite{fievet2020} diagonalize the normal flux derivative and
obtain the characteristic system
\begin{align}
  \Jac\,\dd{\bW}{t} + \cN &= -\cS,
  \label{eq:tp-master}\\
  \cN &= \PU\Big(\dd{\hE}{\xi} - \cA(\xi)\Big),
  \label{eq:tp-N}\\
  \cS &= \PU\Big(\dd{\hF}{\eta} + \dd{\hG}{\zeta} + \cA(\xi)\Big),
  \label{eq:tp-S}
\end{align}
written here with the $\Jac$-weighted convention. The characteristic
variables satisfy $\partial\bW=\PU\,\partial\bU$, where $\PU$ transforms
conservative variables into characteristic ones for the unit normal
$\nabla\xi/\lVert\nabla\xi\rVert$ \cite{fievet2020,marchal2023}. The term
$\cA(\xi)$ gathers the products of the fluxes with the $\xi$-derivatives of
the metric terms. The expressions of $\PU$ and $\cA(\xi)$, which the
simplex treatment does not use, are not repeated here and can be found in
\cite[Sec.~2.2 and Appendix~A]{fievet2020}, and in
\cite[Appendix~1]{marchal2023} for multi-species gases. The normal and
tangential strengths $\cN$ and $\cS$ are built from the full flux $\hE$, so
that they include both the convective and the diffusive contributions,
which is the compact character of the formalism.

A boundary condition replaces some components of $\cN$ by prescribed values,
which defines $\cN^*$. Inverting Eq.~\eqref{eq:tp-N} gives the corrected
flux derivative at the NSCBC face FP,
\begin{equation}
  \Big(\dd{\hE}{\xi}\Big)^{*} = \PU^{-1}\cN^{*} + \cA(\xi).
  \label{eq:tp-dEstar}
\end{equation}

\begin{figure}[htbp]
  \centering
  \includegraphics[width=0.77\linewidth]{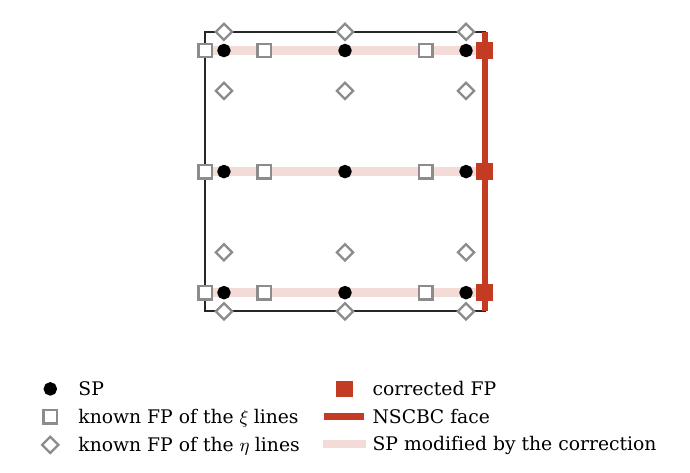}
\caption{Boundary correction on a quadrilateral for $p=2$. Each FP of the
NSCBC face is corrected on its own, and the correction only modifies the
SP of its line, Eqs.~\eqref{eq:tp-Estar} and \eqref{eq:tp-dEsp}.}
\label{fig:coupling-quad}
\end{figure}

The correction then returns to the SP through the line of FP that ends at
the NSCBC face FP $l=p+2$. Since Eq.~\eqref{eq:sd1d-dE} is linear in the FP
values, only $\hE_{p+2}$ needs to change. With $\partial_\xi\Lfp{p+2}$
stored at initialization, Fiévet et al. obtain the corrected boundary flux
and its effect on the SP of that line \cite[Eqs.~(54) and (55)]{fievet2020}:
\begin{align}
  \hE^{*}_{p+2} &= \hE_{p+2}
    + \frac{(\partial_\xi\hE)^{*}_{p+2}-(\partial_\xi\hE)_{p+2}}{\partial_\xi\Lfp{p+2}(1)},
  \label{eq:tp-Estar}\\
  \Big(\dd{\hE}{\xi}\Big)^{*}_{\!j} &= \Big(\dd{\hE}{\xi}\Big)_{\!j}
    + \big(\hE^{*}_{p+2}-\hE_{p+2}\big)\,\partial_\xi\Lfp{p+2}(\xi_j).
  \label{eq:tp-dEsp}
\end{align}
Each NSCBC face FP is thus corrected independently of the others, by a
scalar division per variable. Figure~\ref{fig:coupling-quad} illustrates this
correction on a quadrilateral for $p=2$: each NSCBC face FP only modifies the SP of its line.

This algorithm uses two properties of tensor-product elements. First,
Eqs.~\eqref{eq:tp-N} and \eqref{eq:tp-S} need $\partial_\xi\hE$ separately
from the other terms of the divergence. Second, Eq.~\eqref{eq:tp-Estar} needs
a line of FP normal to the face, along which the correction is
one-dimensional. Section~\ref{sec:sd-simplex} showed that simplex elements
do not have these properties.

%% file: sections/05_nscbc_simplex.tex
\section{NSCBC on simplex elements}
\label{sec:nscbc-simplex}

The extension of the NSCBC treatment to simplex elements is based on three main principles that do not rely on the tensor-product structure. First, the characteristic analysis is performed in the physical frame attached to the boundary face and not in the reference element (see Section~\ref{sec:simplex-char}). Second, the normal derivative of the normal flux is evaluated with the RT divergence operator 
(see Section~\ref{sec:simplex-dn}). Third, the boundary fluxes are recovered through a small linear system local to each boundary element (see Section~\ref{sec:simplex-inverse}).

\subsection{Limitations of the tensor-product formalism with simplex elements}
\label{sec:simplex-why}

Equation~\eqref{eq:tp-N} cannot be evaluated on a simplex element, since
the SDRT operator, Eq.~\eqref{eq:rt-div}, returns the full divergence
$\divxi\hcF$ and no derivative $\partial_\xi\hE$ is available separately. Moreover, on a simplex element a face is not generally at constant reference coordinate: the hypotenuse of the reference triangle is normal to the direction
$(1,1)^\T$ (Fig.~\ref{fig:sp-fp-tri}).

Equation~\eqref{eq:tp-Estar} cannot be used either: it corrects one FP at a
time along a line of FP normal to the face, and such lines do not exist on
simplex elements, where each coefficient of Eq.~\eqref{eq:rt-flux}
contributes to the divergence everywhere in the element.

\begin{figure}[htbp]
  \centering
  \includegraphics[width=0.7\linewidth]{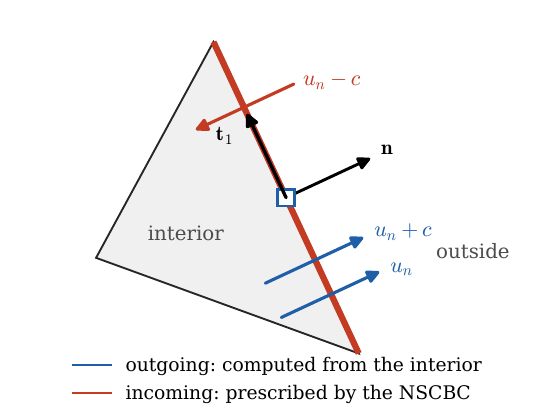}
  \caption{Face frame $(\mathbf{n},\mathbf{t}_1)$ at an FP of an
  inclined NSCBC face, here a subsonic outflow. The waves traveling at
  $u_n$ and $u_n+c$ leave the domain and are computed from the interior. The
  wave traveling at $u_n-c$ enters the domain and is prescribed by the
  boundary condition. 
  }
  \label{fig:face-frame}
\end{figure}

On any element type, each NSCBC face FP has a physical outward unit
normal $\bn$, which defines with two tangential vectors an orthonormal frame
$(\bn,\bt_1,\bt_2)$. Figure~\ref{fig:face-frame} shows this frame at an FP
of an inclined NSCBC face, with the waves that leave and enter the domain
at a subsonic outflow. The whole treatment is built in this frame. The tangential vectors are given in
\ref{app:tangent}, and the frame is computed once per NSCBC face FP during
the pre-processing.

\subsection{Characteristic analysis in the face frame}
\label{sec:simplex-char}
The characteristic analysis is performed on the Euler fluxes, following the
original approach of Poinsot and Lele \cite[Sec.~2.1]{poinsot1992}. Unlike
in the compact formalism of Section~\ref{sec:nscbc-tensor} \cite{fievet2020},
the diffusive fluxes enter neither the wave strengths nor the transverse
terms, and they are handled at the NSCBC face by viscous conditions
(see Section~\ref{sec:simplex-visc}).

Since the physical divergence is independent of the orthonormal frame, it can be written in the frame of a given NSCBC face FP:
\begin{equation}
  \divx\cF = \dd{\En}{x_n} + \dd{\Et{1}}{x_{t_1}} + \dd{\Et{2}}{x_{t_2}},
  \qquad
  \En = \cF\cdot\bn,\quad \Et{i} = \cF\cdot\bt_i ,
  \label{eq:div-rot}
\end{equation}
where $x_n$, $x_{t_1}$ and $x_{t_2}$ are the coordinates along $\bn$,
$\bt_1$ and $\bt_2$. Rotating the momentum components with the matrix:
\begin{equation}
  \Mrot =
  \begin{pmatrix}
    1 & 0 & 0 & 0 & 0\\
    0 & n_x & n_y & n_z & 0\\
    0 & t_{1x} & t_{1y} & t_{1z} & 0\\
    0 & t_{2x} & t_{2y} & t_{2z} & 0\\
    0 & 0 & 0 & 0 & 1
  \end{pmatrix},
  \qquad \Mrot^{-1} = \Mrot^\T,
  \label{eq:Mrot}
\end{equation}
 gives the face-frame conservative variables:
\begin{equation}
  \tU = \Mrot\bU = \big(\rho,\ \rho u_n,\ \rho u_{t_1},\ \rho u_{t_2},\ \rho E\big)^\T,
  \label{eq:tU}
\end{equation}
with $u_n=\bu\cdot\bn$ and $u_{t_i}=\bu\cdot\bt_i$. Since $\Mrot$ is constant
in time and attached to one FP, multiplying Eq.~\eqref{eq:nse-phys} by
$\Mrot$ gives
\begin{equation}
  \dd{\tU}{t} + \dd{\tE}{x_n} + \dd{\tF}{x_{t_1}} + \dd{\tG}{x_{t_2}} = \mathbf{0},
  \label{eq:nse-face}
\end{equation}
with the face-frame fluxes $\tE = \Mrot\En$, $\tF = \Mrot\Et{1}$ and
$\tG = \Mrot\Et{2}$.

These fluxes have the algebraic form of the Cartesian fluxes evaluated
with the velocities $(u_n,u_{t_1},u_{t_2})$, and their Euler part reads:
\begin{equation}
  \tE = \big(\rho u_n,\ \rho u_n^2+P,\ \rho u_n u_{t_1},\
  \rho u_n u_{t_2},\ (\rho E+P)u_n\big)^\T,
  \label{eq:tE}
\end{equation}
where $\tF$ and $\tG$ follow by exchanging the roles of the velocity components.

The face-frame primitive variables is defined as:
\begin{equation}
  \tQ = \big(\rho,\ u_n,\ u_{t_1},\ u_{t_2},\ P\big)^\T,
  \label{eq:tQ}
\end{equation}
and write Eq.~\eqref{eq:nse-face} with them, keeping the transverse terms in
conservative form:
\begin{equation}
  \dd{\tQ}{t} + \An\dd{\tQ}{x_n}
  + \dd{\tQ}{\tU}\left(\dd{\tF}{x_{t_1}}+\dd{\tG}{x_{t_2}}\right) = \mathbf{0},
  \qquad
  \An = \dd{\tQ}{\tU}\dd{\tE}{\tQ}.
  \label{eq:prim-face}
\end{equation}
The matrix $\An$ is the Euler Jacobian in the direction $\bn$, and is
diagonalizable,
\begin{equation}
  \An = \RAn\,\LamAn\,\LAn,
  \qquad
  \LamAn = \mathrm{diag}\big(u_n,u_n,u_n,u_n+c,u_n-c\big),
  \label{eq:An-diag}
\end{equation}
where $\RAn$ gathers the right eigenvectors and $\LAn=\RAn^{-1}$ the left
ones. \ref{app:matrices} gives $\An$, $\RAn$, $\LAn$ and the passage matrices
$\partial\tQ/\partial\tU$ and $\partial\tU/\partial\tQ$.

Multiplying Eq.~\eqref{eq:prim-face} on the left by $\LAn$ yields the characteristic
form
\begin{equation}
  \dd{\bW}{t} + \cL = -\cT,
  \label{eq:char}
\end{equation}
with the characteristic variables, the normal wave strengths and the transverse terms:
\begin{equation}
  \begin{gathered}
  \partial\bW = \LAn\,\partial\tQ,
  \qquad
  \cL = \LamAn\,\dd{\bW}{x_n},\\
  \cT = \LAn\,\dd{\tQ}{\tU}\left(\dd{\tF}{x_{t_1}}+\dd{\tG}{x_{t_2}}\right).
  \end{gathered}
  \label{eq:LT-def}
\end{equation}
The first three components of $\bW$ are the entropy and vorticity waves,
traveling at $u_n$, and the last two are the acoustic waves, traveling at
$u_n+c$ and $u_n-c$.

The wave strengths only depend on the normal derivative of the normal flux:
multiplying Eq.~\eqref{eq:char} by $(\partial\tU/\partial\tQ)\RAn$ and
identifying the result with Eq.~\eqref{eq:nse-face} gives:
\begin{equation}
  \dd{\tE}{x_n} = \dd{\tU}{\tQ}\,\RAn\,\cL
  \quad\Longleftrightarrow\quad
  \cL = \LAn\,\dd{\tQ}{\tU}\,\dd{\tE}{x_n},
  \label{eq:L-from-flux}
\end{equation}
together with
\begin{equation}
  \dd{\tU}{\tQ}\,\RAn\,\cT = \dd{\tF}{x_{t_1}}+\dd{\tG}{x_{t_2}}.
  \label{eq:T-consistency}
\end{equation}
Equation~\eqref{eq:L-from-flux} plays the role of Eq.~\eqref{eq:tp-N} but
involves neither $\partial_\xi\hE$ nor the metric term $\cA(\xi)$: the
reference element only enters through the evaluation of
$\partial\tE/\partial x_n$. In two dimensions, $\bt_2$ and $u_{t_2}$ vanish
and all the matrices reduce to size four.

\subsection{Normal derivative of the normal flux with the RT operator}
\label{sec:simplex-dn}

The normal derivative in Eq.~\eqref{eq:L-from-flux} is the projection of a
physical gradient on $\bn$, which Eq.~\eqref{eq:grad-from-div} expresses
with a reference divergence that the SDRT scheme evaluates on any simplex. For each component $i$ of the normal flux:
\begin{equation}
  \dd{\tEs_i}{x_n} = \gradx\tEs_i\cdot\bn
  = \left(\frac{1}{\Jac}\,\divxi\!\Big(\Jac\,J^{-\T}\,\tEs_i\Big)\right)\cdot\bn,
  \qquad i\in[1,5].
  \label{eq:dn-cont}
\end{equation}

Equation~\eqref{eq:dn-cont} is evaluated with the RT interpolation of
Section~\ref{sec:sd-simplex}. For a scalar field $\phi$, each row of
$\Jac J^{-\T}\phi$ is a vector field whose coefficient at the degree of
freedom $k$ is its component along $\hn_k$, and the divergence of the
interpolant, evaluated at the FP $\bxi_j$ of the NSCBC face, gives
\begin{equation}
  \dd{\phi}{x_n}(\bxi_j)
  = \frac{1}{\Jac_j}\left[\sum_{k=1}^{\NFP}\DFP_{jk}\,\phi(\bxi_k)\,
    \Jac_k J_k^{-\T}\hn_k\right]\cdot\bn_j,
  \label{eq:dn-disc}
\end{equation}
with the differentiation matrix evaluated at the FP,
\begin{equation}
  \DFP_{jk} = \big(\divxi\bpsi_k\big)(\bxi_j).
  \label{eq:DFP}
\end{equation}
Equation~\eqref{eq:dn-disc} is the gradient formula of Veilleux et al.
\cite[Eq.~(42)]{veilleux2022jcp}, evaluated at an FP instead of an SP.
For a given NSCBC face FP $j$, it is applied to $\phi=\tEs_i$ with the frame of
that point, the fluxes being evaluated at all the FP of the element.

The evaluation of the wave strengths at the NSCBC face FP follows five steps:
\begin{enumerate}[label=\arabic*.,leftmargin=*,nosep]
  \item interpolate the solution at all the FP of the boundary element,
  \item rotate the variables with $\Mrot_j$
        and evaluate $\tE$ at all the NSCBC face FP $j$ of the element,
  \item compute $\partial\tE/\partial x_n$ at $\bxi_j$ with
        Eq.~\eqref{eq:dn-disc},
  \item compute $\cL$ with Eq.~\eqref{eq:L-from-flux},
  \item compute $\cT$ in the same way, with $\bt_1$ and $\bt_2$ in place of
        $\bn$ (see Section~\ref{sec:simplex-transverse}).
\end{enumerate}

\subsection{Recovering the normal flux at the NSCBC face FP}
\label{sec:simplex-inverse}

The boundary condition replaces some components of $\cL$, which defines
$\cL^*$ (see Section~\ref{sec:simplex-bc}). At the NSCBC face FP $j$,
inverting Eq.~\eqref{eq:L-from-flux} gives
$(\partial\tE/\partial x_n)^{*} = (\partial\tU/\partial\tQ)\,\RAn\,\cL^{*}$,
the corrected normal derivative of the face-frame flux.
The face-frame flux is $\tE = \Mrot_j\En$, where $\Mrot_j$ is built with the
normal and tangents of FP $j$ and is used unchanged at every point of the
element (see Section~\ref{sec:simplex-dn}). This frozen matrix commutes with
$\partial/\partial x_n$, so that
$\partial\tE/\partial x_n = \Mrot_j\,\partial\En/\partial x_n$, and
multiplying by $\Mrot_j^{-1}=\Mrot_j^\T$ gives the corrected normal
derivative in the Cartesian frame,
\begin{equation}
  \Big(\dd{\En}{x_n}\Big)^{*}_{j} = \Mrot_j^\T\,\dd{\tU}{\tQ}\,\RAn\,\cL^{*}.
  \label{eq:dEn-star}
\end{equation}
In this expression, $\En$ stands for $\cF\cdot\bn_j$ at every point of the
element, including the other FP of the NSCBC face.
This is the counterpart of Eq.~\eqref{eq:tp-dEstar}, and it remains to find
the boundary fluxes that produce this derivative.

The imposed normal fluxes are then deduced by inverting
Eq.~\eqref{eq:dn-disc}. Its right-hand
side involves all the FP of the element, including the FP of the NSCBC face
$\dK$ where the normal flux is now unknown, and moving the known values to
the right-hand side gives, for every FP $j\in\dK$ and every component $i$,
\begin{equation}
  \sum_{k'\in\dK} B_{jk'}\,E_{n,i}^{*}(\bxi_{k'})
  = \Jac_j\Big(\dd{E_{n,i}}{x_n}\Big)^{*}_{j}
  - \sum_{k\notin\dK} B_{jk}\,E_{n,i}(\bxi_k),
  \label{eq:inv-system}
\end{equation}
with the geometric coefficients
\begin{equation}
  B_{jk} = \DFP_{jk}\,\big(\Jac_k J_k^{-\T}\hn_k\big)\cdot\bn_j .
  \label{eq:Bjk}
\end{equation}
Equation~\eqref{eq:inv-system} is a dense linear system whose size is the
number of FP on the face,
\begin{equation}
  \NFPb = p+1\ \text{(2D)},
  \qquad
  \NFPb = \frac{(p+1)(p+2)}{2}\ \text{(3D)}.
  \label{eq:nfpb}
\end{equation}
The matrix $B_{jk'}$ is the same for the five components and depends on the
geometry only, so that it can be computed and inverted once per boundary
element during the pre-processing.

\begin{figure}[htbp]
  \centering
  \includegraphics[width=0.77\linewidth]{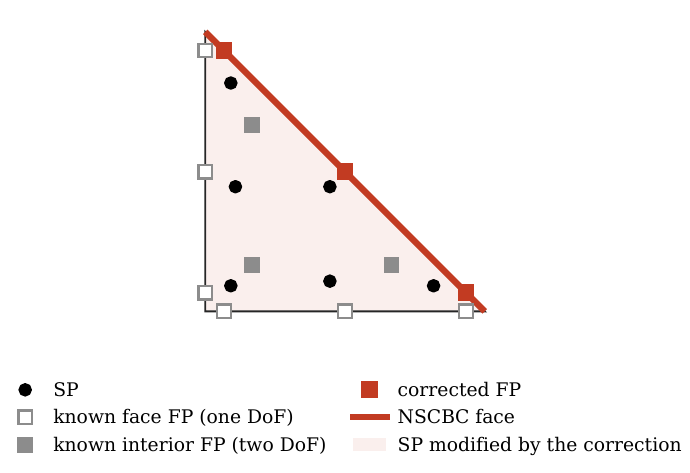}
  \caption{Boundary correction on a triangle for $p=2$. The FP of the NSCBC
  face are the unknowns of the linear system Eq.~\eqref{eq:inv-system},
  whose right-hand side involves all the other FP, and the correction
  modifies every SP of the element.}
  \label{fig:coupling-tri}
\end{figure}

The solution of the linear system is the convective numerical flux at each
NSCBC face FP. On a face FP, the
vector $\Jac_k J_k^{-\T}\hn_k$ is parallel to the physical normal,
$\Jac_k J_k^{-\T}\hn_k = \lVert\Jac_k J_k^{-\T}\hn_k\rVert\,\bn_k$, so that
the coefficients of Eq.~\eqref{eq:rt-coef} on $\dK$ become
\begin{equation}
  \big(\hcF_{k'}\cdot\hn_{k'}\big)^{*} = \lVert\Jac_{k'} J_{k'}^{-\T}\hn_{k'}\rVert\,\En^{*}(\bxi_{k'}),
  \qquad k'\in\dK ,
  \label{eq:bnd-coef}
\end{equation}
and Eq.~\eqref{eq:rt-div} advances the boundary element in time.
Equations~\eqref{eq:inv-system} and \eqref{eq:bnd-coef} replace
Eq.~\eqref{eq:tp-Estar}, and the correction is no longer a scalar update
per FP: the NSCBC face FP are coupled through one system, fed by every FP
of the element. Figure~\ref{fig:coupling-tri} illustrates this coupling on a
triangle for $p=2$, where the correction modifies every SP of the
element.

Equation~\eqref{eq:inv-system} relies on one assumption on the geometry of
the NSCBC face. The equation written at the NSCBC face FP $j$ is built in
the frame of $j$, so that the flux at every other NSCBC face FP $k'$ enters
it projected on $\bn_j$, as $\cF(\bxi_{k'})\cdot\bn_j$. The scheme, however,
only needs $\cF(\bxi_{k'})\cdot\bn_{k'}$ at that point,
Eq.~\eqref{eq:bnd-coef}. Solving for one unknown per NSCBC face FP
identifies these two quantities, which is exact on a planar face, where all
the normals coincide. All the NSCBC faces of Section~\ref{sec:results} are
planar, and the treatment of curved NSCBC faces is left for future work.
\subsection{Transverse terms}
\label{sec:simplex-transverse}

The transverse terms use the same principle without any linear system: for
each NSCBC face FP, $\partial\tF/\partial x_{t_1}$ and
$\partial\tG/\partial x_{t_2}$ are evaluated with Eq.~\eqref{eq:dn-disc},
replacing $\bn_j$ by $\bt_{1,j}$ and $\bt_{2,j}$, and
Eq.~\eqref{eq:LT-def} then gives $\cT$. These terms are not modified by the
boundary condition, and enter the corrected wave strengths through the
inflow and outflow relations below.

\subsection{Subsonic inflow and outflow}
\label{sec:simplex-bc}

With $\bn$ pointing outwards, the characteristic form Eq.~\eqref{eq:char}
reads in primitive variables
\begin{equation}
  \dd{}{t}
  \begin{pmatrix} \rho\\ u_n\\ u_{t_1}\\ u_{t_2}\\ P \end{pmatrix}
  = -
  \begin{pmatrix}
    \cL_1+\cT_1+\frac{\rho}{\sqrt{2}c}\,\Sigma_{45}\\[2pt]
    \frac{1}{\sqrt{2}}\big(\cL_4-\cL_5+\cT_4-\cT_5\big)\\[2pt]
    -\frac{c}{\rho}\big(\cL_3+\cT_3\big)\\[2pt]
    \frac{c}{\rho}\big(\cL_2+\cT_2\big)\\[2pt]
    \frac{\rho c}{\sqrt{2}}\,\Sigma_{45}
  \end{pmatrix},
  \label{eq:prim-explicit}
\end{equation}
where $\Sigma_{45} = \cL_4+\cL_5+\cT_4+\cT_5$.
A boundary condition prescribes the time derivatives of some primitive
variables, each prescription determining one incoming wave strength through
Eq.~\eqref{eq:prim-explicit}, while the outgoing ones keep their interior
values.

\paragraph{Subsonic inflow} At a subsonic inflow, $u_n<0$, $u_n+c>0$ and
$u_n-c<0$, so that four waves enter the domain and $\cL_1$, $\cL_2$,
$\cL_3$ and $\cL_5$ are unknown. The velocity and the temperature must
remain at targets $\bu^\tgt$ and $T^\tgt$ that are constant in time. The
exact conditions $\partial_t\bu=\mathbf{0}$ and $\partial_t T=0$ would keep
the boundary values constant, but would not bring them back to the targets
once numerical errors have moved them away, and the inflow values would
drift \cite[Sec.~3.1.2]{fievet2020}. Following \cite{fievet2020}, these
time derivatives are therefore replaced by linear relaxation terms,
$\partial_t\bu=K_u(\bu^\tgt-\bu)$ and $\partial_t T=K_T(T^\tgt-T)$, with the
relaxation coefficients $K_u$ and $K_T$, and the velocity components give
\begin{align}
  \cL^*_5 &= \cL_4+\cT_4-\cT_5+\sqrt{2}\,K_u\big(\bu^\tgt-\bu\big)\cdot\bn,
  \label{eq:in-un}\\
  \cL^*_3 &= -\cT_3+\frac{\rho}{c}\,K_u\big(\bu^\tgt-\bu\big)\cdot\bt_1,
  \label{eq:in-ut1}\\
  \cL^*_2 &= -\cT_2-\frac{\rho}{c}\,K_u\big(\bu^\tgt-\bu\big)\cdot\bt_2 .
  \label{eq:in-ut2}
\end{align}
Differentiating the equation of state $T=P/(\rho\Rgas)$ in time and
inserting Eq.~\eqref{eq:prim-explicit} gives the last strength,
\begin{equation}
  \cL^*_1 = -\cT_1+\frac{\rho}{P}\left[\rho\Rgas K_T\big(T^\tgt-T\big)
    +\frac{\cL_4+\cL^*_5+\cT_4+\cT_5}{\sqrt{2}}\Big(\rho c-\frac{P}{c}\Big)\right].
  \label{eq:in-L1}
\end{equation}

\paragraph{Subsonic outflow} At a subsonic outflow, $u_n>0$, $u_n+c>0$ and
$u_n-c<0$, so that only the acoustic wave $\cL_5$ enters the domain
(Fig.~\ref{fig:face-frame}). The pressure must remain at a target
$P^\tgt$ constant in time, but the exact condition $\partial_t P=0$ lets the
mean pressure drift, since the domain receives no pressure feedback from
the outside \cite{poinsot1992,fievet2020}. As proposed by Poinsot and Lele
\cite{poinsot1992}, the pressure is therefore relaxed towards $P^\tgt$
with the coefficient $K_P$. Prescribing $\partial_t P = K_P(P^\tgt-P)$ in
Eq.~\eqref{eq:prim-explicit} would give
\begin{equation}
  \cL^*_5 = -\frac{\sqrt{2}\,K_P}{\rho c}\big(P^\tgt-P\big)-\cL_4-\cT_4-\cT_5,
  \label{eq:out-refl}
\end{equation}
a condition that reflects the outgoing acoustic wave $\cL_4$. We
therefore remove $\cL_4$ and $\cT_4$. Following \cite{yoo2007,lodato2008}, we also weight the
transverse terms by $(1-\alpha)$:
\begin{equation}
  \cL^*_5 = -\frac{\sqrt{2}\,K_P}{\rho c}\big(P^\tgt-P\big) - (1-\alpha)\,\cT_5,
  \label{eq:out}
\end{equation}
where $\alpha$ is usually set to the mean Mach number on the outlet
\cite{yoo2007}.

\paragraph{Reduced relaxation coefficients} The coefficients $K_u$,
$K_T$ and $K_P$ are rates, in $\si{s^{-1}}$, and their effect depends on
the size of the domain. For a subsonic outflow, Rudy and
Strikwerda~\cite{rudy1980} showed that the optimal relaxation
coefficient is proportional to the speed of sound divided by the length
of the domain.
Following Daviller et al.~\cite{daviller2019}, we characterize each case
by the reduced coefficients
\begin{equation}
  \sigma_u=\frac{K_u\Lref}{c},\qquad
  \sigma_T=\frac{K_T\Lref}{c},\qquad
  \sigma_P=\frac{K_P\Lref}{c},
  \label{eq:sigma}
\end{equation}
where $\Lref$ is the length of the domain along the mean flow and $c$
the speed of sound of the target state. Each reduced coefficient compares
the acoustic time $\Lref/c$, needed by a wave to cross the domain, with
the relaxation time $1/K$. These coefficients refer to the
relaxation terms of Eqs.~\eqref{eq:in-un} to \eqref{eq:out}, whereas the
tensor-product formulation multiplies its relaxation terms by
$\Jac$~\cite[Eqs.~(30) and (32)]{fievet2020}, so that the same $K$ does not
give the same relaxation there.

\subsection{Diffusive flux at the NSCBC face}
\label{sec:simplex-visc}

The characteristic treatment only provides the convective part of the
boundary flux, and the diffusive part follows the viscous conditions of Poinsot
and Lele \cite{poinsot1992}, which depend on the type of boundary. With the
diffusive flux of Eq.~\eqref{eq:fluxes}, the normal diffusive flux at an NSCBC
face FP reads
\begin{equation}
  \cF_d\cdot\bn = \big(0,\ -(\bm{\tau}\bn)^\T,\ \mathbf{q}\cdot\bn-(\bm{\tau}\bn)\cdot\bu\big)^\T,
  \label{eq:Fdn}
\end{equation}
and only the normal stress and heat flux, $\bm{\tau}\bn$ and
$\mathbf{q}\cdot\bn$, need to be specified.

At a subsonic outflow, the normal derivatives of the tangential viscous
stresses and of the normal heat flux vanish \cite[Table~IV]{poinsot1992}:
\begin{equation}
  \dd{}{x_n}\big[(\bm{\tau}\bn)\cdot\bt_1\big] = 0,
  \qquad
  \dd{}{x_n}\big[(\bm{\tau}\bn)\cdot\bt_2\big] = 0,
  \qquad
  \dd{}{x_n}\big(\mathbf{q}\cdot\bn\big) = 0 .
  \label{eq:visc-out}
\end{equation}
These conditions are imposed with the linear system of
Section~\ref{sec:simplex-inverse}: each quantity of Eq.~\eqref{eq:visc-out}
is evaluated at all the FP of the element from the interior solution and
its gradient, and Eq.~\eqref{eq:inv-system}, written with a zero target
derivative, then gives its values at the NSCBC face FP. The same matrix
$B_{jk'}$ thus serves for the convective and the diffusive corrections.
The normal component $(\bm{\tau}\bn)\cdot\bn$ of the normal stress keeps
its interior value, and
Eq.~\eqref{eq:Fdn} assembles from these values the normal diffusive flux,
which is added to the convective flux of Eq.~\eqref{eq:bnd-coef}.

At a subsonic inflow, the velocity and the temperature are imposed, which
is case SI~1 of Poinsot and Lele \cite[Table~III]{poinsot1992}: the density
follows from the continuity equation, which contains no viscous term, and
no viscous condition is required. The diffusive flux at an NSCBC face FP is therefore evaluated with Eq.~\eqref{eq:Fdn} from the solution and the gradients of the boundary element at that point, without involving the neighboring element as at an interior interface.

\subsection{Algorithm and comparison with tensor-product elements}
\label{sec:simplex-algo}

At each stage of the time integration, every boundary element with an
NSCBC face undergoes the following steps:
\begin{enumerate}[label=\arabic*.,leftmargin=*,nosep]
  \item Evaluate $\cL$ and $\cT$ at each NSCBC face FP
        (see Section~\ref{sec:simplex-dn}),
  \item Build $\cL^*$ from the boundary condition
        (see Section~\ref{sec:simplex-bc}),
  \item Compute $(\partial\En/\partial x_n)^*$ with
        Eq.~\eqref{eq:dEn-star},
  \item Solve Eq.~\eqref{eq:inv-system} for the normal fluxes on the face,
  \item Replace the convective numerical flux with Eq.~\eqref{eq:bnd-coef},
  \item Add the diffusive flux of Section~\ref{sec:simplex-visc} and
        advance the element with Eq.~\eqref{eq:rt-div}.
\end{enumerate}

Table~\ref{tab:comparison} sums up the differences with the tensor-product
treatment of Section~\ref{sec:nscbc-tensor}.

\begin{table}[htbp]
  \centering
  \caption{Tensor-product and simplex NSCBC treatments.}
  \label{tab:comparison}
  \small
  \begin{tabular}{@{}>{\raggedright\arraybackslash}p{0.24\linewidth}>{\raggedright\arraybackslash}p{0.34\linewidth}>{\raggedright\arraybackslash}p{0.34\linewidth}@{}}
    \toprule
    & Tensor-product \cite{fievet2020} & Simplex (this work) \\
    \midrule
    Frame of the analysis & reference element, face $\xi=1$ & physical face frame $(\bn,\bt_1,\bt_2)$ \\
    Normal & $\nabla\xi/\lVert\nabla\xi\rVert$ & $\bn$ of the FP \\
    Wave strengths & $\cN$, Eq.~\eqref{eq:tp-N} & $\cL$, Eq.~\eqref{eq:L-from-flux} \\
    Transverse terms & $\cS$, Eq.~\eqref{eq:tp-S} & $\cT$, Eq.~\eqref{eq:LT-def} \\
    Fluxes in the strengths & convective and diffusive & convective only \\
    Metric term & $\cA(\xi)$ & none \\
    Normal derivative & $\partial_\xi\hE$ along a line of FP, Eq.~\eqref{eq:sd1d-dE} & $\partial\tE/\partial x_n$ with the RT divergence, Eq.~\eqref{eq:dn-disc} \\
    Corrected quantity & $\hE_{p+2}$, full interface flux & $\En$, convective numerical flux \\
    Diffusive face flux & in $\hE_{p+2}$ & viscous conditions, Section~\ref{sec:simplex-visc} \\
    Feedback & scalar update per FP, Eq.~\eqref{eq:tp-Estar} & linear system of size $\NFPb$, Eq.~\eqref{eq:inv-system} \\
    SP affected & SP of the line of the FP & all SP of the element \\
    \bottomrule
  \end{tabular}
\end{table}

%% file: sections/06_results.tex
\section{Validation test cases}
\label{sec:results}

Unless stated otherwise, the numerical parameters used in the test cases below are as follows. The convective numerical fluxes come from the Harten-Lax-van Leer-Contact (HLLC) Riemann solver of Batten et al.~\cite{batten1997}, and time integration uses the three-stage, third-order strong-stability-preserving Runge-Kutta (RK) scheme of Gottlieb and Shu~\cite{gottlieb1998} with a constant time step $\Delta t$ given for each case. The SDRT scheme (see Section~\ref{sec:sd-simplex}) is used on triangles, and C-SDRT on tetrahedra, since no linearly stable SDRT formulation has been found on tetrahedra for $p>2$~\cite{veilleux2022jsc}. Only the inflows and outflows use the NSCBC treatment of Section~\ref{sec:nscbc-simplex}, and the other boundaries of the test cases, walls and symmetry planes, are not treated with characteristic conditions. All results will be compared to the tensor-product formulation of Fi\'evet et al.~\cite{fievet2020}, which serves as a baseline, for every tests cases using the same polynomial degree and almost the same number of DoF.

\subsection{Isentropic vortex through a straight outlet}
\label{sec:res-vortex}
The objective of this case is to check that the transverse terms let a vortex leave the domain without reflection, and that the simplex treatment reproduces the tensor-product one. Following the
configuration of Fi\'evet et al.~\cite{fievet2020}, the Euler equations are solved in the square $[0,\SI{0.1}{m}]^2$, where a uniform flow at
$M_\infty=0.3$ ($u_\infty=\SI{104}{m/s}$), $T_\infty=\SI{300}{K}$ and
$P_\infty=\SI{e5}{Pa}$ carries an isentropic vortex centered at
$(x_c,y_c)=(\SI{0.05}{m},\SI{0.05}{m})$. With
$d^2=(x-x_c)^2+(y-y_c)^2$, the initial state reads
\begin{equation}
  \left\{
  \begin{aligned}
    u &= u_\infty-\Gamma\,\frac{y-y_c}{R_v}\,e^{\displaystyle-d^2/(2R_v^2)},\\[1ex]
    v &= \Gamma\,\frac{x-x_c}{R_v}\,e^{\displaystyle-d^2/(2R_v^2)},\\[1ex]
    T &= T_\infty-\frac{\Gamma^2}{2C_p}\,e^{\displaystyle-d^2/R_v^2},
  \end{aligned}
  \right.
  \label{eq:vortex-init}
\end{equation}
with $\rho=\rho_\infty(T/T_\infty)^{1/(\gamma-1)}$, $R_v=\SI{5}{mm}$ and
$\Gamma=\SI{34.7}{m/s}$.

The right boundary is the outlet under study and uses
Eq.~\eqref{eq:out} with $K_P=0$ and $\alpha=0.3$, so that only the
transverse terms prevent the reflection. The top and bottom boundaries
are outlets without transverse terms, and the left boundary is a
subsonic inflow that imposes $u_\infty$, $v=0$ and $T_\infty$
(see Section~\ref{sec:simplex-bc}). Three configurations are compared: triangles with and without transverse terms at the right outlet, and quadrilaterals with transverse terms, which serve as the reference~\cite{fievet2020}.

\begin{figure}[htbp]
  \centering
  \includegraphics[width=\linewidth]{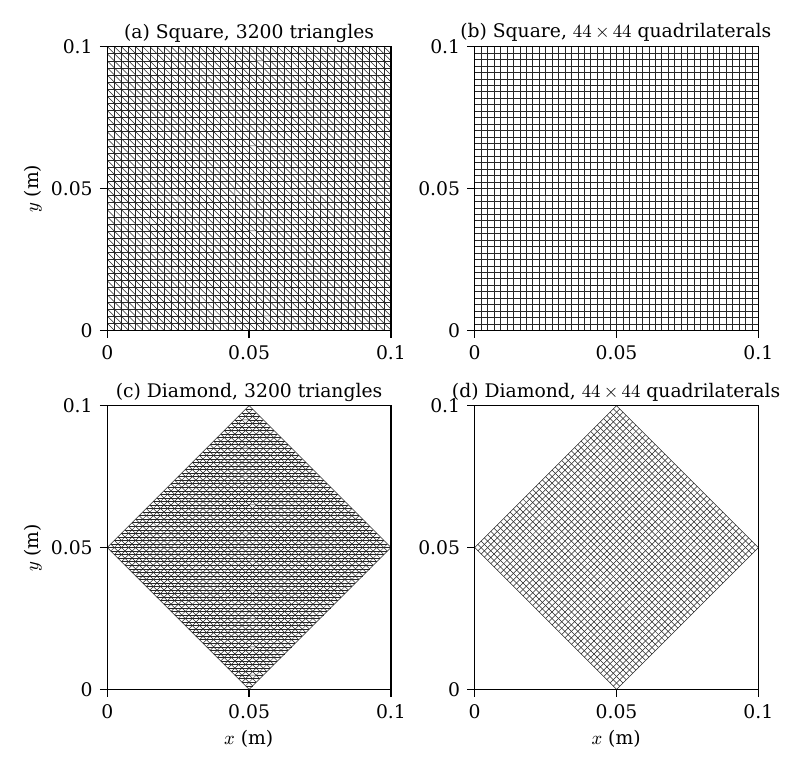}
  \caption{Meshes of the isentropic vortex cases, all structured.
  Straight outlet (see Section~\ref{sec:res-vortex}): (a) triangles,
  (b) quadrilaterals. Inclined outlet
  (see Section~\ref{sec:res-vortex-incl}): (c) triangles,
  (d) quadrilaterals.}
  \label{fig:vortex-meshes}
\end{figure}

Figures~\ref{fig:vortex-meshes}a and~\ref{fig:vortex-meshes}b show the
two meshes, both structured, which give the same resolution at $p=4$: 3200 triangles carry \num{48000} SP and
$44\times44$ quadrilaterals carry \num{48400} SP. The runs use
$\Delta t=\SI{5e-8}{s}$ up to $t=\SI{1.2}{ms}$.

Following \cite{fievet2020}, the pressure is normalized as
$P^*=(P-P_\infty)/P_0$, where $P_0=P_\infty-\min P$ is taken at $t=0$
($P_0\approx\SI{697}{Pa}$). Two metrics, computed with the extrema of
$P^*$ over all SP, quantify the reflection: the amplitude
$P^*_{\max}-P^*_{\min}$ and the drift $P^*_{\max}+P^*_{\min}$. Both are
meaningful only once the vortex core has left the domain, its center
reaching the outlet at $t\approx\SI{0.48}{ms}$.

\begin{figure}[htbp]
  \centering
  \includegraphics[width=\linewidth]{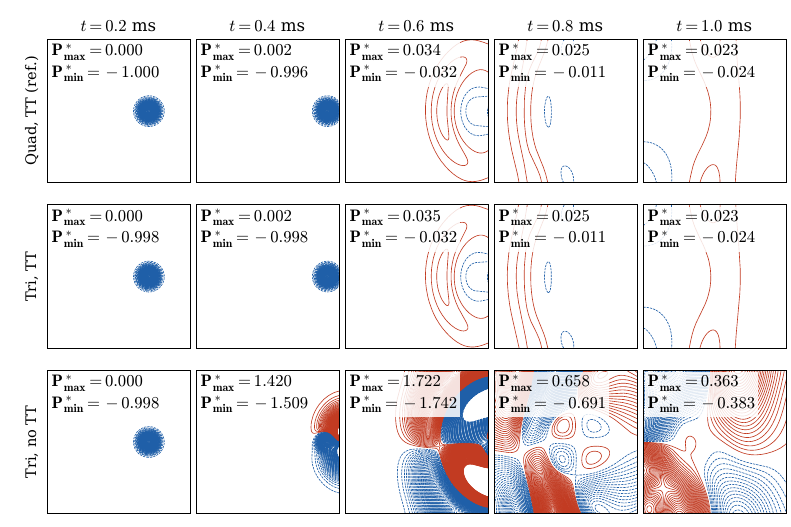}
  \caption{Vortex through a straight outlet: isolines of $P^*$ every
  $1\,\%$ (red: positive, blue: negative), at several times. Top:
  quadrilaterals with transverse terms (TT,
  reference). Middle: triangles with TT. Bottom: triangles without TT.
  The extrema of $P^*$ are given in each frame.}
  \label{fig:vortex-square-p}
\end{figure}

Figure~\ref{fig:vortex-square-p} shows that the triangles with
transverse terms and the quadrilaterals cannot be told apart. At
$t=\SI{0.4}{ms}$, the vortex touches the outlet and $P^*_{\max}=0.002$
in both runs, whereas it already reaches $1.42$ without transverse
terms. At $t=\SI{0.6}{ms}$, a weak reflected wave travels upstream,
with $P^*_{\max}=0.0345$ and $P^*_{\min}=-0.0316$ in both runs, whereas
without transverse terms the same wave reaches $P^*_{\max}=1.72$ and
$P^*_{\min}=-1.74$ and the reflection is larger than the vortex itself.

\begin{figure}[htbp]
  \centering
  \includegraphics[width=\linewidth]{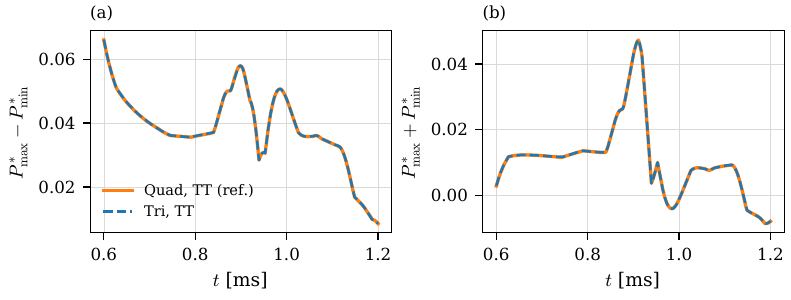}
  \caption{Vortex through a straight outlet, runs with transverse terms
  (TT): (a) reflection amplitude and (b) pressure drift after the vortex
  has left.}
  \label{fig:vortex-square-metrics}
\end{figure}

\begin{figure}[htbp]
  \centering
  \includegraphics[width=\linewidth]{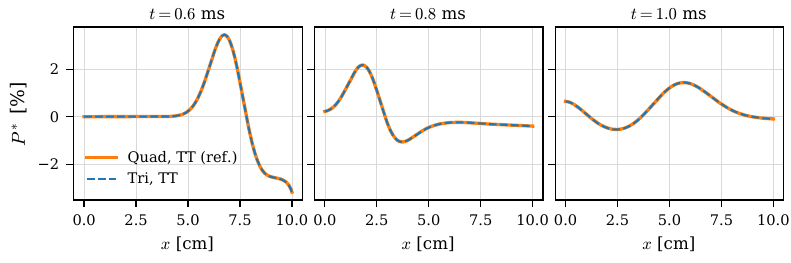}
  \caption{Vortex through a straight outlet, runs with transverse terms
  (TT): $P^*$ along $y=y_c$ at three times.}
  \label{fig:vortex-square-cut}
\end{figure}

Figure~\ref{fig:vortex-square-metrics} shows the reflection amplitude
and the drift once the vortex core has left the domain, and confirms the
visual agreement. At $t=\SI{0.6}{ms}$, the amplitude is $0.0662$ on triangles and $0.0661$
on quadrilaterals, against $3.46$ without transverse terms. The two
curves with transverse terms remain superimposed until the end of the
run, including after $t\approx\SI{0.85}{ms}$, when both metrics
oscillate. This time matches the arrival of the reflected waves at the
inflow, since they travel upstream at $c-u_\infty\approx\SI{243}{m/s}$
and need about $\SI{0.4}{ms}$ to cross the domain. At $t=\SI{1.2}{ms}$,
the drift is $-0.0078$ in both runs. Figure~\ref{fig:vortex-square-cut}
shows the profiles of $P^*$ along $y=y_c$, which also coincide: on this
case, the simplex treatment reproduces the tensor-product one at equal
resolution.

\subsection{Isentropic vortex through an inclined outlet}
\label{sec:res-vortex-incl}
This case checks that the treatment stays non-reflecting when the outlet
is inclined with respect to the flow, the situation for which
Section~\ref{sec:nscbc-simplex} builds the face frame. The flow and the vortex of Section~\ref{sec:res-vortex} are the same, Eq.~\eqref{eq:vortex-init}, but the domain is the square rotated by $\SI{45}{\degree}$, with vertices
$(0.05,0)$, $(0.1,0.05)$, $(0.05,0.1)$ and $(0,0.05)$ in meters. The
vortex starts at $(x_c,y_c)=(\SI{0.05}{m},\SI{0.07}{m})$ and travels
along $x$, reaching the upper right face, $x+y=\SI{0.15}{m}$, at
$t\approx\SI{0.29}{ms}$ and crossing it at $\SI{45}{\degree}$.

The two right faces are outlets that use Eq.~\eqref{eq:out} with
$K_P=0$ and $\alpha=0.3$, with or without transverse terms, and the two
left faces are subsonic inflows that impose $u_\infty$, $v=0$ and
$T_\infty$, that is, the uniform flow of the initial state. 

Figures~\ref{fig:vortex-meshes}c and~\ref{fig:vortex-meshes}d show the
two meshes, both structured, which give the same resolution at $p=4$:
3200 triangles carry \num{48000} SP and $44\times44$ quadrilaterals carry
\num{48400} SP. The $\Delta t$ of
Section~\ref{sec:res-vortex} is kept, up to $t=\SI{0.8}{ms}$.

\begin{figure}[htbp]
  \centering
  \includegraphics[width=\linewidth]{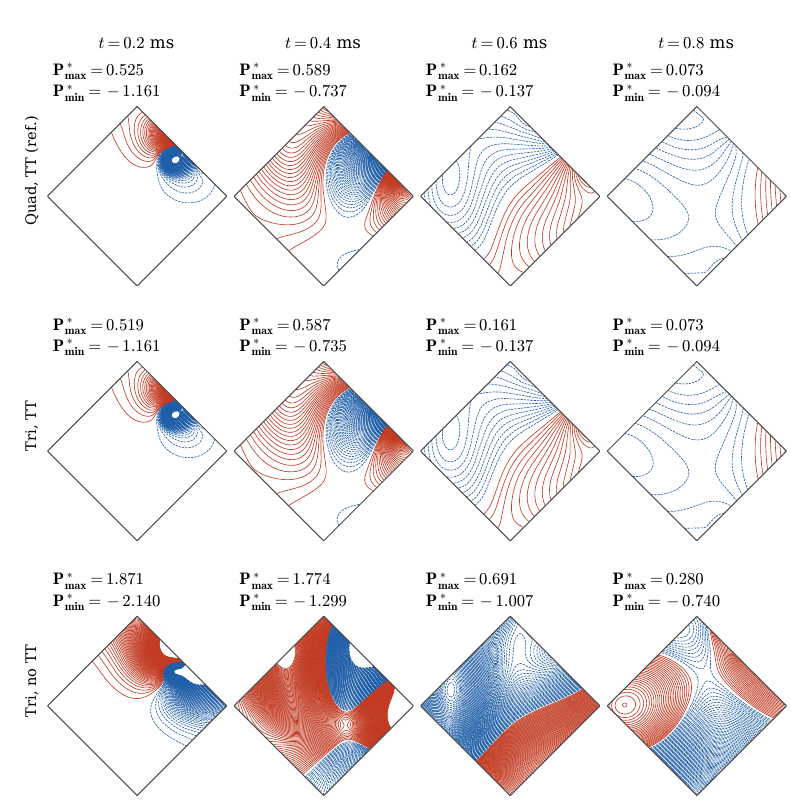}
  \caption{Vortex through an inclined outlet: isolines of $P^*$ every
  $1\,\%$ (red: positive, blue: negative), at several times. Top:
  quadrilaterals with transverse terms (TT,
  reference). Middle: triangles with TT. Bottom: triangles without TT.
  The outlets are the two right faces.}
  \label{fig:vortex-diamond-p}
\end{figure}

Figure~\ref{fig:vortex-diamond-p} shows that the triangles with
transverse terms again reproduce the quadrilaterals. When the vortex
crosses the inclined face, the pressure drops well below the vortex
core value: $P^*_{\min}$ reaches $-2.77$ on triangles and $-2.79$ on
quadrilaterals, at $t\approx\SI{0.28}{ms}$, and since both
discretizations show this drop, it does not come from the simplex
treatment. Once the vortex has left, the remaining waves are weak and
coincide in both runs, whereas without transverse terms the pressure
field is strongly perturbed during the whole run.

\begin{figure}[htbp]
  \centering
  \includegraphics[width=\linewidth]{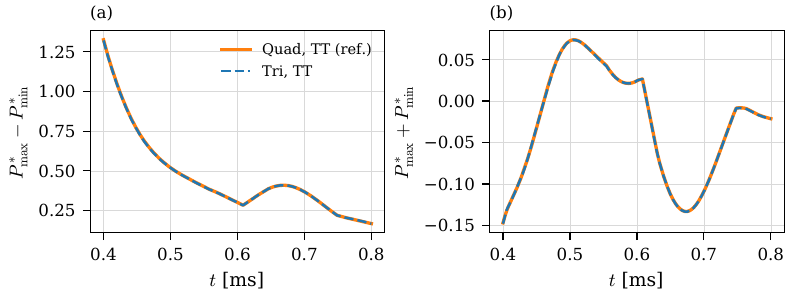}
  \caption{Vortex through an inclined outlet, runs with transverse terms
  (TT): (a) reflection amplitude and (b) pressure drift.}
  \label{fig:vortex-diamond-metrics}
\end{figure}

Figure~\ref{fig:vortex-diamond-metrics} shows the reflection amplitude
and the drift for the inclined outlet, and confirms the agreement. After $t=\SI{0.4}{ms}$, the amplitude and the drift on triangles remain
very close to those on quadrilaterals, and differ by at most 0.006 and 0.002, respectively. At $t=\SI{0.8}{ms}$, the amplitude
is $0.167$ on triangles and $0.168$ on
quadrilaterals, against $1.02$ without transverse terms, and the drift
is $-0.021$ in both runs, against $-0.46$ without transverse terms. The
simplex NSCBC treatment is thus as effective on an inclined outlet as the
tensor-product one.

\subsection{Planar Poiseuille flow}
\label{sec:res-poiseuille}
This case checks that the inflow and outflow conditions impose a target
velocity profile and a target pressure in a viscous flow. The
Navier-Stokes equations are solved in the channel
$[0,\SI{3}{m}]\times[-H/2,H/2]$, with $H=\SI{1}{m}$, a constant dynamic
viscosity $\mu=\SI{0.05}{Pa.s}$ and $\mathrm{Pr}=0.72$, between adiabatic
no-slip walls at $y=\pm H/2$. The inflow at $x=0$ relaxes the velocity towards
the parabolic profile
\begin{equation}
  u^\tgt(y)=U_{\max}\left(1-\frac{4y^2}{H^2}\right),\qquad v^\tgt=0,
  \label{eq:poiseuille-u}
\end{equation}
with $U_{\max}=\SI{4}{m/s}$, and the temperature towards
$T^\tgt=\SI{300}{K}$, with $K_u=K_T$. The outflow at $x=\SI{3}{m}$
uses Eq.~\eqref{eq:out} with $P^\tgt=\SI{101325}{Pa}$ and
$\alpha=0.00768$ (the bulk Mach number). Two sets of runs are considered. In the first, $K_P$ is fixed at $\SI{1000}{s^{-1}}$ and $K_u$ is varied over 100, 1000 and $\SI{e4}{s^{-1}}$; in the second, $K_u$ is fixed at $\SI{1000}{s^{-1}}$ and $K_P$ is varied over 10, 100, 1000 and $\SI{e4}{s^{-1}}$.
On triangles, with $\Lref=\SI{3}{m}$ and $c=\SI{347}{m/s}$, these values
give $\sigma_u=\sigma_T=0.86$, $8.6$ and $86$, and $\sigma_P=0.086$,
$0.86$, $8.6$ and $86$ (see Eq.~\eqref{eq:sigma}).
The flow starts at rest, at $P^\tgt$ and $T^\tgt$, and the Reynolds number
based on $H$ and on the bulk velocity $2U_{\max}/3$ is about 63.

The steady solution is the plane Poiseuille flow: $u=u^\tgt(y)$ in the
whole channel and a uniform pressure gradient:
\begin{equation}
  \frac{\mathrm{d}P}{\mathrm{d}x}=-\frac{8\mu U_{\max}}{H^2}
  =\SI{-1.6}{Pa/m},
  \label{eq:poiseuille-dpdx}
\end{equation}
that is, a pressure drop of $\SI{4.8}{Pa}$ between the inflow and the
outflow.

\begin{figure}[htbp]
  \centering
  \includegraphics[width=\linewidth]{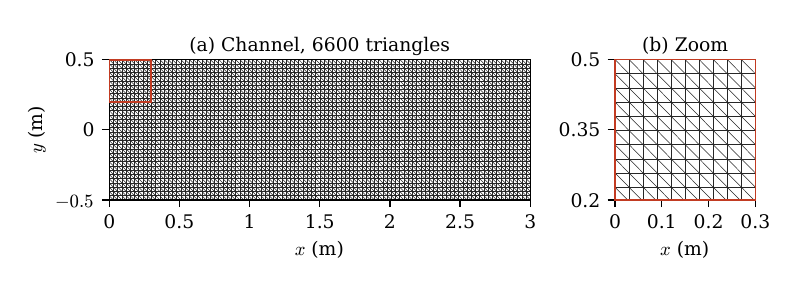}
  \caption{Poiseuille flow: triangular mesh. (a) Whole channel, (b) zoom
  on the region framed in red, at the inflow and the upper wall.}
  \label{fig:poiseuille-mesh}
\end{figure}

Both meshes use $p=2$ and carry \num{39600} SP: 6600 triangles, obtained
by splitting each cell of a $100\times33$ structured grid into two
triangles, and $100\times44$ quadrilaterals. Figure~\ref{fig:poiseuille-mesh}
shows the triangular mesh, with a zoom at the inflow near the upper wall.
The runs use $\Delta t=\SI{6.1944e-6}{s}$ and stop after $10^6$
iterations, at $t=\SI{6.19}{s}$. This time corresponds to about 5.5
flow-through times based on the bulk velocity, and all runs are steady
well before it. Two probes, at the SP
closest to $(0,0)$ and to $(\SI{3}{m},0)$, monitor the solution at the
inflow and at the outflow, respectively.

\begin{figure}[htbp]
  \centering
  \includegraphics[width=\linewidth]{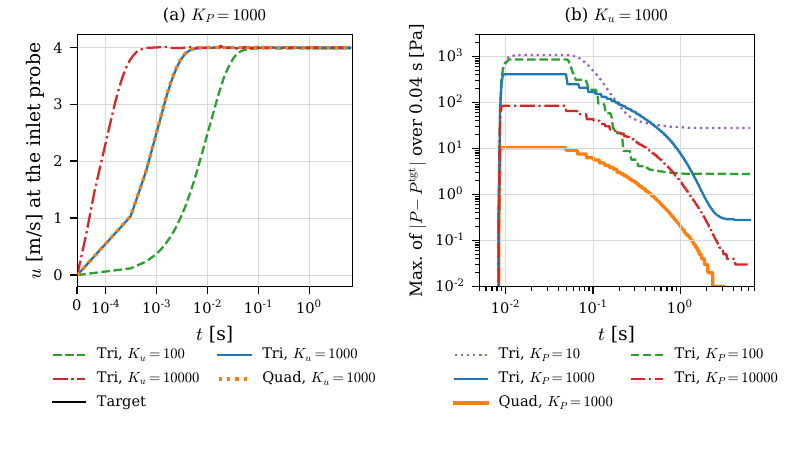}
  \caption{Poiseuille flow: (a) streamwise velocity at the inlet probe
  for several $K_u$, the black line marking the target (time axis
  linear below $\SI{e-4}{s}$, logarithmic above), (b) maximum of
  $|P-P^\tgt|$ at the outlet probe over a sliding time window, for
  several $K_P$, the quadrilateral curve leaving the axis when the
  departure falls below the resolution of the probe.}
  \label{fig:poiseuille-probes}
\end{figure}

Figure~\ref{fig:poiseuille-probes}a shows the streamwise velocity at the
inlet probe: on both meshes and for all $K_u$, the inflow velocity reaches
its target. The time to come within
$1\,\%$ of the target decreases as $K_u$ increases.

The outflow pressure takes longer to settle, as acoustic waves make it
oscillate during the transient. Figure~\ref{fig:poiseuille-probes}b
therefore shows the maximum of $|P-P^\tgt|$ over the previous $\SI{0.04}{s}$,
which filters out these oscillations. All runs are steady
to within $\SI{0.01}{Pa}$ before $t=\SI{3.4}{s}$.
On quadrilaterals with $K_P=1000$, the final outlet pressure matches
$P^\tgt$ to within the $\SI{0.01}{Pa}$ resolution of the probe, whereas
on triangles it stays above $P^\tgt$ by \SI{27.86}{Pa}, \SI{2.79}{Pa},
\SI{0.28}{Pa} and \SI{0.03}{Pa} for $K_P=10$, $100$, $1000$ and $10^4$.
This departure decreases as $1/K_P$ and remains below $\num{3e-6}\,P^\tgt$
for $K_P\geq1000$.

\begin{figure}[htbp]
  \centering
  \includegraphics[width=\linewidth]{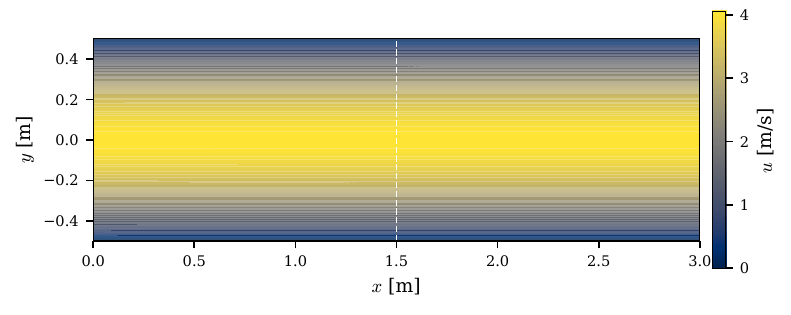}
  \caption{Poiseuille flow on triangles, $K_u=1000$ and $K_P=10^4$: streamwise
  velocity at $t=\SI{6.19}{s}$. The white dashed line marks the cut of
  Fig.~\ref{fig:poiseuille-cuts}a.}
  \label{fig:poiseuille-map}
\end{figure}

\begin{figure}[htbp]
  \centering
  \includegraphics[width=\linewidth]{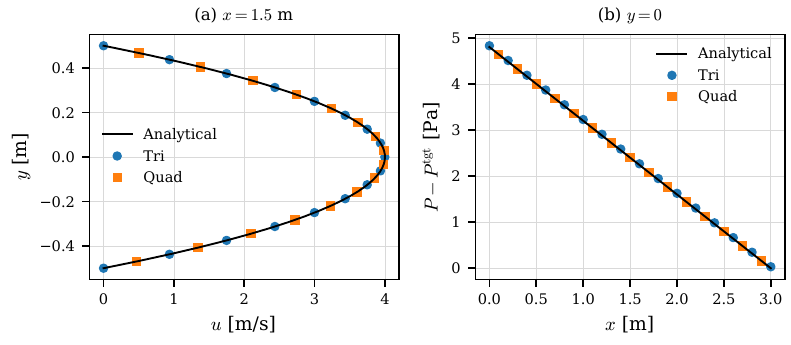}
  \caption{Poiseuille flow, $K_u=1000$, $t=\SI{6.19}{s}$, with
  $K_P=10^4$ on triangles and $K_P=1000$ on quadrilaterals:
  (a) velocity profile across the channel, (b) pressure along the
  centerline, against the analytical solution.}
  \label{fig:poiseuille-cuts}
\end{figure}

Figure~\ref{fig:poiseuille-map} shows the steady streamwise velocity on
triangles, which is the expected Poiseuille flow, and
Figure~\ref{fig:poiseuille-cuts} compares both discretizations with the
analytical solution. Both use $K_u=1000$, and $K_P$ is set to $10^4$ on
triangles and to $1000$ on quadrilaterals, so that the outlet pressure
of both runs departs from $P^\tgt$ by at most $\SI{0.03}{Pa}$ at the probe. On both meshes, the
velocity profiles at $x=\SI{1.5}{m}$ coincide with
Eq.~\eqref{eq:poiseuille-u}, and the pressure along $y=0$ varies
linearly with the gradient of Eq.~\eqref{eq:poiseuille-dpdx}.

The triangles and the quadrilaterals therefore give close results, and
the small differences between them most likely come from the slight
differences between the two formulations
(see Sections~\ref{sec:nscbc-tensor} and~\ref{sec:nscbc-simplex}).

\subsection{Spherical pressure wave in a cube}
\label{sec:res-sphere}

This case checks that the outlets let an acoustic wave leave the domain
on tetrahedra, when the wave reaches the boundary under oblique
incidence. Navier-Stokes equations are solved in a cube of side $\Lref=\SI{13}{mm}$ whose six
faces are outlets~\cite{lodato2008}. The fluid is initially at rest, at
$P_\infty=\SI{101325}{Pa}$ and $T_\infty=\SI{300}{K}$, and a Gaussian
pressure pulse is placed at the center of the cube. With $d$ the
distance to the center, the initial pressure reads
\begin{equation}
  P = P_\infty\left[1+\varepsilon\,e^{\displaystyle-d^2/(2R_s^2)}\right],
  \label{eq:sphere-init}
\end{equation}
with $\varepsilon=10^{-3}$ and $R_s=0.05\,\Lref$, and the density
follows from the equation of state at the uniform temperature
$T_\infty$. The pulse generates a spherical acoustic wave that reaches
the centers of the faces first, then the edges and the corners. The
outlets therefore see angles of incidence from zero, at the center of a
face, up to $\arccos(1/\sqrt{3})\approx\ang{55}$, at a corner.

The reflections are measured with the residual fluctuating pressure of
Bogey and Bailly~\cite{Bogey2002e},
\begin{equation}
  L_P(t)=\left[\frac{1}{N}\sum_{i=1}^{N}\big(P_i(t)-P_\infty\big)^2\right]^{1/2},
  \label{eq:sphere-lp}
\end{equation}
where the sum runs over the $N$ SP of the mesh. While the wave is inside the
domain, $L_P$ follows the free-field solution. Once the wave has left,
$L_P$ only measures the waves reflected by the outlets and the
numerical errors, and an ideal outlet gives $L_P=0$. The analytical
curve applies Eq.~\eqref{eq:sphere-lp} to the free-field solution of a
Gaussian pulse given by Bogey and Bailly~\cite{Bogey2002e}. As in their
work, time is normalized by $t_0=\Delta/c_0$, with $\Delta=\Lref/128$
and $c_0=\SI{343}{m/s}$, which gives $t_0=\SI{2.96e-7}{s}$.

\begin{figure}[htbp]
  \centering
  \includegraphics[width=\linewidth]{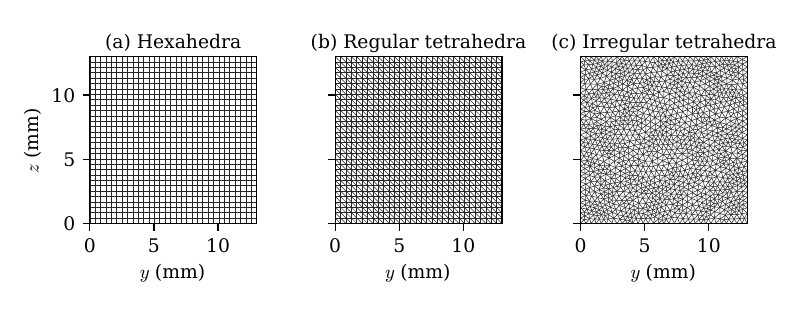}
  \caption{Spherical pressure wave in a cube: face $x=0$ of (a) the
  hexahedral mesh, (b) the regular tetrahedral mesh and (c) the
  irregular tetrahedral mesh.}
  \label{fig:sphere-meshes}
\end{figure}

We compare the three meshes of Fig.~\ref{fig:sphere-meshes}. The
hexahedral mesh is a regular grid of $31^3=\num{29791}$ hexahedra. The
regular tetrahedral mesh splits each of these hexahedra into six
tetrahedra, which gives \num{178746} tetrahedra. The irregular
tetrahedral mesh is generated by the Delaunay algorithm of Gmsh with a
uniform characteristic length $\Lref/32$ and has \num{132624}
tetrahedra. At $p=3$, the hexahedral mesh carries \num{1906624} SP, and
the regular and irregular tetrahedral meshes carry \num{3574920} and
\num{2652480} SP, respectively. The time step is set by a Courant-Friedrichs-Lewy (CFL)
number of 0.2 on hexahedra and 0.1 on tetrahedra. This gives $\Delta t=\SI{6.04e-8}{s}$ for the hexahedral mesh; $\SI{3.02e-8}{s}$ and $\SI{1.58e-8}{s}$ for the regular and irregular tetrahedral meshes, respectively. The runs stop at $t=250\,t_0$. The outlets use Eq.~\eqref{eq:out} with
$P^\tgt=P_\infty$, $K_P=\SI{10}{s^{-1}}$ and $\alpha=10^{-3}$. With
$\Lref$ the side of the cube, this gives $\sigma_P=3.7\times10^{-4}$
(see Eq.~\eqref{eq:sigma}). Each mesh is run with and without
transverse terms, which gives six runs.

\begin{figure}[htbp]
  \centering
  \includegraphics[width=\linewidth]{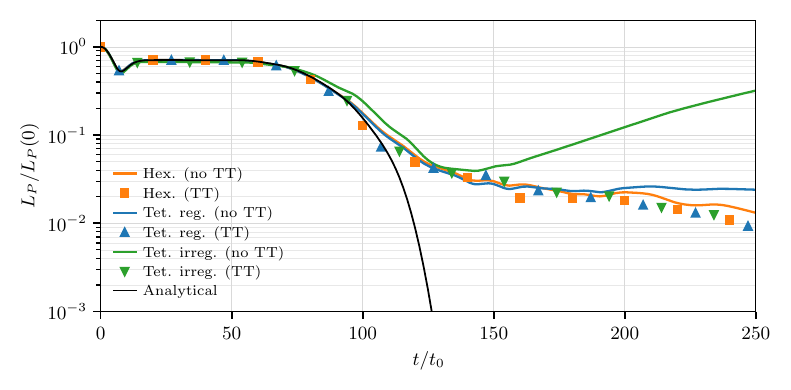}
  \caption{Spherical pressure wave in a cube: time evolution of the
  residual fluctuating pressure $L_P$ (see Eq.~\eqref{eq:sphere-lp}),
  normalized by its initial value, on the three meshes with (TT) and
  without (no TT) transverse terms, and analytical solution.}
  \label{fig:sphere-lp}
\end{figure}

Figure~\ref{fig:sphere-lp} shows the time evolution of $L_P$ for the
six runs. Until about $t/t_0=110$, the runs with transverse terms follow
the analytical curve on the three meshes. The analytical curve then
falls below $10^{-3}$ at about $t/t_0=125$, when the wave has left the
cube, whereas the computed values remain of the order of $10^{-2}$ or
above. These residual values measure the waves reflected by the outlets.

With transverse terms, the three meshes give superimposed curves, and
$L_P/L_P(0)$ decreases down to about $10^{-2}$ at $t/t_0=250$
(Fig.~\ref{fig:sphere-lp}). The residual pressure thus ends two orders
of magnitude below its initial value, the level that Bogey and
Bailly~\cite{Bogey2002e} obtain with their far-field boundary
conditions on a Cartesian grid. The irregular tetrahedral mesh, whose
faces are not aligned with the wave fronts, does not degrade this
result.

Without transverse terms, the result depends on the mesh. On
hexahedra, $L_P$ stays close to the runs with transverse terms. On the
regular tetrahedral mesh, $L_P/L_P(0)$ levels off at about
$2\times10^{-2}$ after $t/t_0=150$. On the irregular tetrahedral mesh,
$L_P$ leaves the analytical curve first, near $t/t_0=85$, and grows
after $t/t_0=130$, up to about $0.3\,L_P(0)$ at $t/t_0=250$. On
tetrahedra, the transverse terms are therefore needed to let the wave
leave the domain, as for the vortex of Section~\ref{sec:res-vortex}.

\subsection{Poiseuille flow in a pipe}
\label{sec:res-3d}
This case checks the formulation on tetrahedra with curved wall faces, in
a viscous flow where the inflow sets the flow rate and the outflow sets
the pressure level. The Navier-Stokes equations are solved in a pipe of
radius $R_p=\SI{0.5}{mm}$ and length $\SI{5}{mm}$ along $x$, with a
constant dynamic viscosity $\mu=\SI{1.81e-5}{Pa.s}$, $\mathrm{Pr}=0.72$
and an adiabatic no-slip wall. The inflow at $x=0$ relaxes the
velocity towards the profile
\begin{equation}
  u^\tgt(y,z)=U_{\max}\left(1-\frac{y^2+z^2}{R_p^2}\right),\qquad
  v^\tgt=w^\tgt=0,
  \label{eq:pipe-u}
\end{equation}
with $U_{\max}=\SI{15}{m/s}$, and the temperature towards
$T^\tgt=\SI{300}{K}$. The outflow at $x=\SI{5}{mm}$ uses
Eq.~\eqref{eq:out} with $P^\tgt=\SI{101325}{Pa}$ and $\alpha=0.0216$
(the bulk Mach number). The flow starts at rest, at $P^\tgt$ and
$T^\tgt$, and the Reynolds number based on the diameter and on the bulk
velocity $U_{\max}/2$ is about 490.

The steady solution is the Hagen-Poiseuille flow: $u=u^\tgt(y,z)$ in the
whole pipe and a uniform pressure gradient
\begin{equation}
  \frac{\mathrm{d}P}{\mathrm{d}x}=-\frac{4\mu U_{\max}}{R_p^2}
  =\SI{-4344}{Pa/m},
  \label{eq:pipe-dpdx}
\end{equation}
that is, a pressure drop of $\SI{21.7}{Pa}$ along the pipe.

\begin{figure}[htbp]
  \centering
  \includegraphics[width=\linewidth]{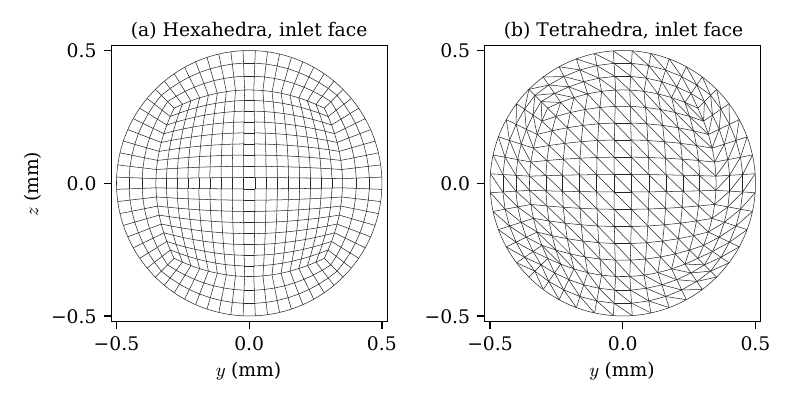}
  \caption{Poiseuille flow in a pipe: inlet face of (a) the hexahedral
  mesh and (b) the tetrahedral mesh.}
  \label{fig:pipe-meshes}
\end{figure}

Two second-order meshes are compared, with a curved wall and 25 element
layers along $x$, whose inlet faces are shown in Fig.~\ref{fig:pipe-meshes}. The tetrahedral mesh has
\num{37950} elements and the hexahedral mesh \num{12325}, which carry
\num{759000} and \num{788800} SP at $p=3$. In this test case, the time integration uses the five-stage, fourth-order
strong-stability-preserving RK scheme of Spiteri and
Ruuth~\cite{spiteri2002}. The time step varies to keep a
CFL number of 0.15 on tetrahedra and 0.35 on
hexahedra. 
Two probes, at the SP closest to $(0,0,0)$ and to $(\SI{5}{mm},0,0)$,
monitor the solution at the inflow and at the outflow, respectively. The results on the tetrahedral and hexahedral meshes are compared at $t=\SI{1.3}{ms}$ and $t=\SI{2}{ms}$, respectively. %

On tetrahedra, $K_u=K_T=\SI{e5}{s^{-1}}$ and $K_P$ is set to $10^3$,
$10^4$ or $\SI{e5}{s^{-1}}$, which give $\sigma_u=\sigma_T=1.4$ and
$\sigma_P=0.014$, $0.14$ and $1.4$ with $\Lref=\SI{5}{mm}$ and
$c=\SI{347}{m/s}$ (see Eq.~\eqref{eq:sigma}). The hexahedral run uses
$K_u=K_T=K_P=\SI{e4}{s^{-1}}$.

\begin{figure}[htbp]
  \centering
  \includegraphics[width=\linewidth]{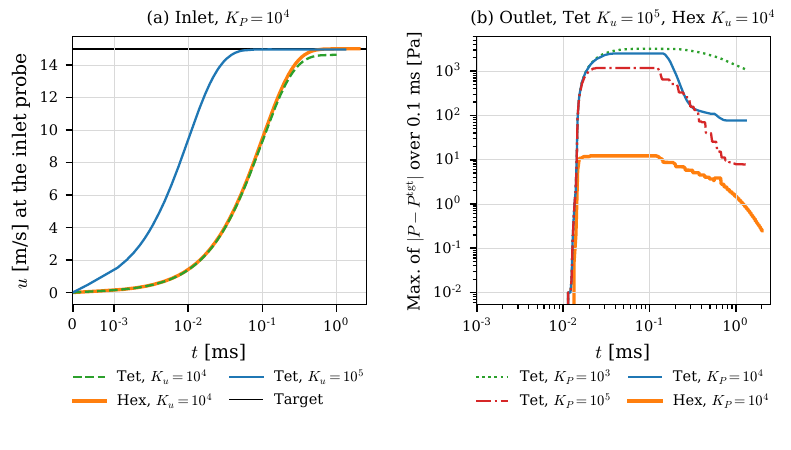}
  \caption{Poiseuille flow in a pipe: (a) streamwise velocity at the
  inlet probe for several $K_u$, the black line marking the target at
  the tetrahedral probe (time axis linear below $\SI{e-3}{ms}$,
  logarithmic above), (b) maximum of $|P-P^\tgt|$ at the outlet probe
  over a sliding time window, which filters out the acoustic
  oscillations, for several $K_P$.}
  \label{fig:pipe-probes}
\end{figure}

Figure~\ref{fig:pipe-probes}a shows the streamwise velocity at the inlet
probe: on both meshes, the inflow velocity reaches its target. On tetrahedra, a larger $K_u$ improves the convergence of the inflow velocity towards its target, the residual offset being roughly inversely proportional to the relaxation coefficient: with $K_u=K_T=K_P=\SI{e4}{s^{-1}}$, a tetrahedral run stopped at $t=\SI{1}{ms}$ levels off $\SI{0.37}{m/s}$ below the target, and multiplying $K_u$ by a decade reduces this deficit by a factor of ten.

On tetrahedra, the outflow pressure settles above $P^\tgt$
(Fig.~\ref{fig:pipe-probes}b). After $t=\SI{1.0}{ms}$, the departure
stays at $\SI{76.5}{Pa}$ with $K_P=10^4$ and at $\SI{7.7}{Pa}$ with
$K_P=10^5$, so that it decreases as $1/K_P$, as on triangles
(see Section~\ref{sec:res-poiseuille}). With $K_P=10^3$, the relaxation
time $1/K_P=\SI{1}{ms}$ is of the order of the simulated time, and the
departure still decreases at $t=\SI{1.3}{ms}$, where it reaches
$\SI{1036}{Pa}$. On hexahedra, the outlet pressure ends $\SI{0.05}{Pa}$
above $P^\tgt$.

\begin{figure}[htbp]
  \centering
  \includegraphics[width=\linewidth]{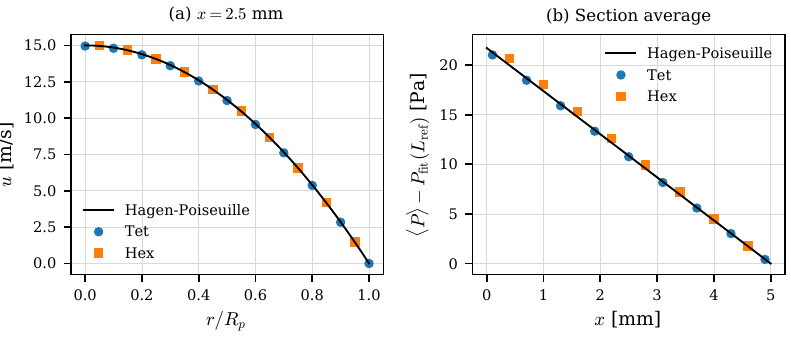}
  \caption{Poiseuille flow in a pipe, tetrahedra with
  $K_u=\SI{e5}{s^{-1}}$ and $K_P=\SI{e4}{s^{-1}}$ at $t=\SI{1.3}{ms}$,
  hexahedra at $t=\SI{2}{ms}$: (a) velocity profile at mid-length,
  averaged over eight azimuths, (b) pressure averaged over 49
  cross-sections between $x=\SI{0.1}{mm}$ and $\SI{4.9}{mm}$, shifted by the value $P_{\mathrm{fit}}(\Lref)$ of
  its linear fit at $x=\Lref$, against Eqs.~\eqref{eq:pipe-u}
  and~\eqref{eq:pipe-dpdx}.}
  \label{fig:pipe-profiles}
\end{figure}

Figure~\ref{fig:pipe-profiles} compares the tetrahedral run with
$K_P=10^4$ and the hexahedral run with the Hagen-Poiseuille flow. At
$x=\SI{2.5}{mm}$, the velocity profiles of both meshes are almost
identical and lie on Eq.~\eqref{eq:pipe-u}. The pressure averaged over
cross-sections, $\langle P\rangle$, varies linearly with $x$. Its
gradient is $\SI{-4297}{Pa/m}$ on tetrahedra, very close to
Eq.~\eqref{eq:pipe-dpdx}, and $\SI{-4503}{Pa/m}$ on hexahedra, that is,
$3.7\,\%$ above it in magnitude. With $K_P=10^5$, the tetrahedral
gradient is $\SI{-4293}{Pa/m}$: the outlet departure shifts the pressure
level in the pipe without changing its gradient.

On tetrahedra with curved wall faces, the simplex formulation therefore
behaves as on triangles. The inflow imposes the velocity profile, the
outflow sets the pressure level with a departure that decreases as
$K_P$ increases, and the pressure gradient in the pipe is very close to
the Hagen-Poiseuille value.

\subsection{Application: two-dimensional premixed burner}
\label{sec:res-burner}
This case applies the formulation to a reacting flow on triangles, in
which a premixed flame stabilizes at the exit of a slot injector: the inflow
feeds fresh gases a few millimeters upstream of the flame, and the
outflow lets the burnt gases leave the domain. 
\ref{app:multispecies} extends the formulation to multi-species
reacting flows: governing equations, numerical ingredients and
characteristic boundary conditions.

\begin{figure}[htbp]
  \centering
  \input{figures/burner_sketch.tex}
  \caption{Two-dimensional premixed burner: domain and boundary
  conditions, after Marchal et al.~\cite[Fig.~12]{marchal2023}. The
  vertical scale is stretched by a factor of 2.3. The flame is drawn
  schematically.}
  \label{fig:burner-sketch}
\end{figure}
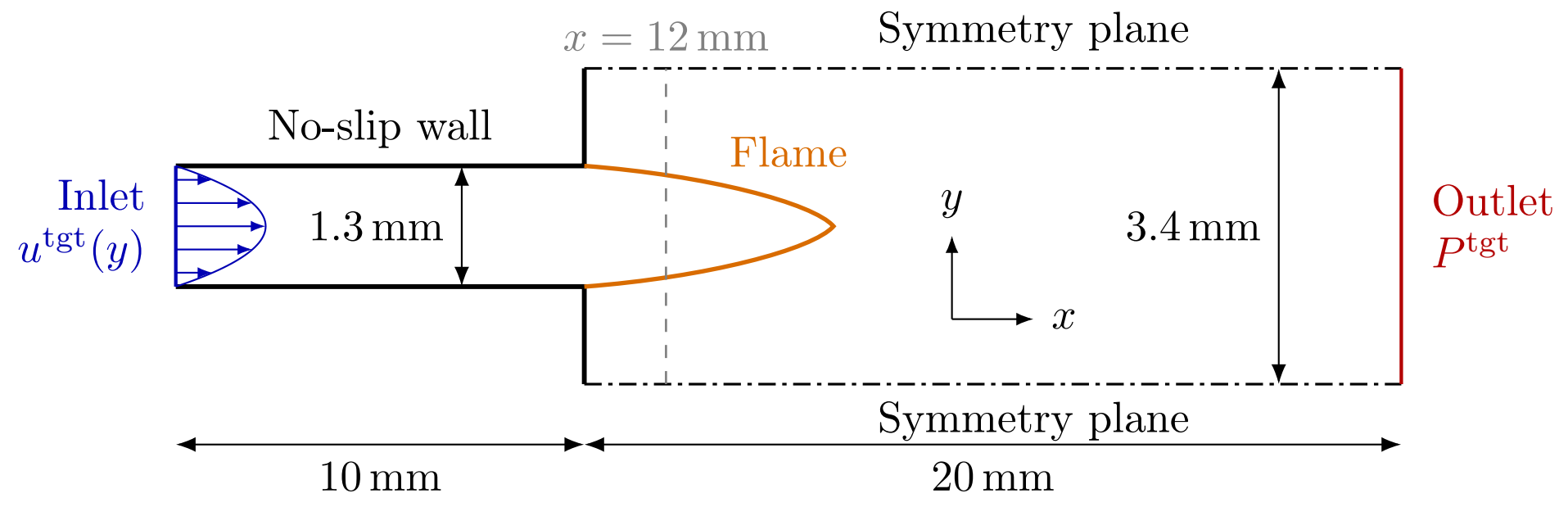

The geometry and inflow condition are defined in Marchal et
al.~\cite[Appendix~3]{marchal2023}.  Fig.~\ref{fig:burner-sketch} shows
the domain and the boundary conditions. The injector is a channel of length
$\SI{10}{mm}$ and width $\SI{1.3}{mm}$ that opens at $x=\SI{10}{mm}$ into
a chamber of length $\SI{20}{mm}$ and width $\SI{3.4}{mm}$, so that
$\Lref=\SI{30}{mm}$. The injector walls and the step at
$x=\SI{10}{mm}$ are adiabatic no-slip walls, and the lateral boundaries
of the chamber are symmetry planes. A methane-air mixture (equivalence ratio of 0.8) enters at $x=0$, where the inflow relaxes the
velocity towards the profile:
\begin{equation}
  u^\tgt(y)=u_0\left(1-\frac{y^2}{l_0^2}\right),\qquad v^\tgt=0,
  \label{eq:burner-u}
\end{equation}
with $u_0=\SI{4}{m/s}$ and $l_0=\SI{0.65}{mm}$, the temperature towards
$T^\tgt=\SI{300}{K}$, and the mass fractions towards
$Y_{\mathrm{CH_4}}^\tgt=0.0446$, $Y_{\mathrm{O_2}}^\tgt=0.2226$ and
$Y_{\mathrm{N_2}}^\tgt=0.7328$ (see Eq.~\eqref{eq:ms-inY}), with
$K_u=K_T=K_Y=\SI{e5}{s^{-1}}$. The outflow at $x=\SI{30}{mm}$ uses
Eq.~\eqref{eq:out} with $P^\tgt=\SI{101325}{Pa}$,
$K_P=\SI{e3}{s^{-1}}$ and $\alpha=10^{-3}$.
We take $c=\SI{354}{m/s}$, the speed of sound of the fresh gases at
$\SI{300}{K}$ with a heat capacity ratio of 1.4, close to that of air.
With $\Lref=\SI{30}{mm}$, these values give $\sigma_u=\sigma_T=8.5$ and
$\sigma_P=0.085$ (see Eq.~\eqref{eq:sigma}).
The chemistry follows the two-step BFER scheme of Franzelli et
al.~\cite{franzelli2010}, with the transport model of
\ref{app:multispecies}.

Both meshes use $p=4$: the triangular mesh has \num{2052} elements and
\num{30780} SP, and the quadrilateral mesh has \num{1216} elements and \num{30400} SP (Marchal et
al.~\cite{marchal2023}). The time step varies
to keep a CFL number of 0.15, and triangles use the C-SDRT formulation(see Section~\ref{sec:sd-simplex}). Both runs
start from a planar flame at the injector exit: fresh gases at
$\SI{300}{K}$ upstream, burnt gases at $\SI{2010}{K}$ downstream, and a
uniform velocity of $\SI{0.282}{m/s}$, and stop at $t=\SI{15}{ms}$.

\begin{figure}[htbp]
  \centering
  \includegraphics[width=\linewidth]{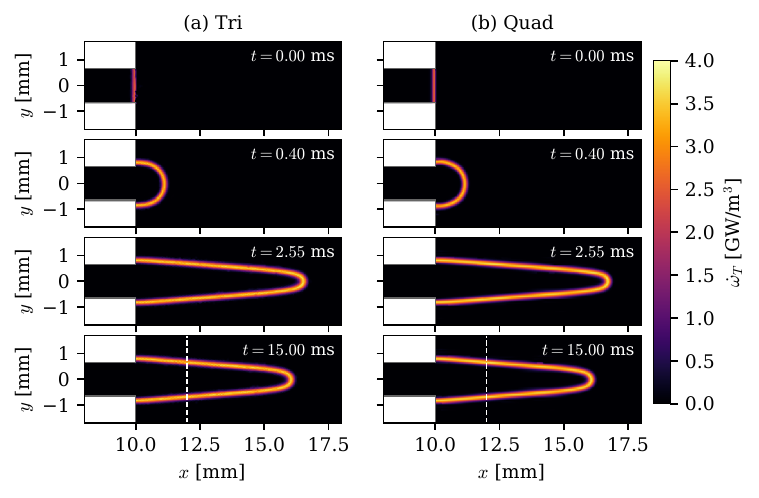}
  \caption{Two-dimensional premixed burner: heat release rate
  $\dot{\omega}_T$ on (a) triangles and (b) quadrilaterals at several
  times. The white dashed line marks the cut $x=\SI{12}{mm}$ of
  Fig.~\ref{fig:burner-cut}.}
  \label{fig:burner-snapshots}
\end{figure}

\begin{figure}[htbp]
  \centering
  \includegraphics[width=\linewidth]{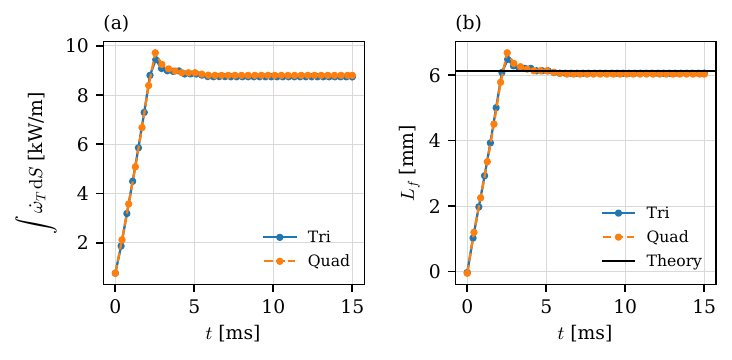}
  \caption{Two-dimensional premixed burner: (a) heat release rate
  integrated over the domain, per unit depth, (b) flame length $L_f$,
  against Eq.~\eqref{eq:burner-lf}. Markers show the outputs.}
  \label{fig:burner-history}
\end{figure}

Figure~\ref{fig:burner-snapshots} shows the heat release rate at four
instants on both meshes. The flame first bulges into the chamber, then stretches into a V-shaped flame anchored at the injector lips.
The flame length $L_f$ is defined as the distance between the injector exit and the maximum of $\dot{\omega}_T$ along the horizontal axis $y=0$.
Figure~\ref{fig:burner-history} shows the heat release rate integrated
over the domain and $L_f$ as functions of time. Both grow until
$t\approx\SI{2.5}{ms}$, overshoot, and decrease towards their final
values, which they reach on both meshes from $t\approx\SI{6.3}{ms}$ on.

\begin{figure}[htbp]
  \centering
  \includegraphics[width=\linewidth]{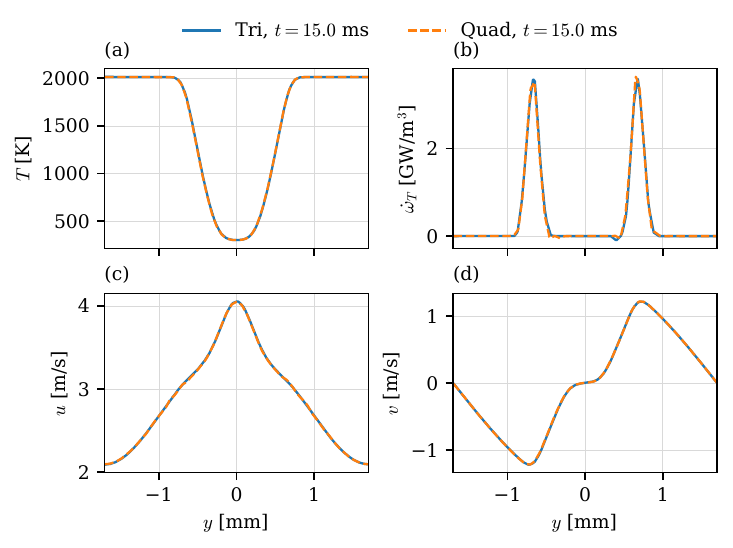}
  \caption{Two-dimensional premixed burner, profiles along
  $x=\SI{12}{mm}$ at the end of the runs: (a) temperature, (b) heat release rate,
  (c) streamwise and (d) transverse velocity.}
  \label{fig:burner-cut}
\end{figure}

Figure~\ref{fig:burner-cut} shows the profiles across the flame at
$x=\SI{12}{mm}$ and $t=\SI{15}{ms}$, which coincide on both meshes. The maximum
temperature is $\SI{2012.7}{K}$ on triangles and $\SI{2013.5}{K}$ on
quadrilaterals. The peak of $\dot{\omega}_T$ is
$\SI{3.594}{GW/m^3}$ at $y=\pm\SI{0.680}{mm}$ on triangles and
$\SI{3.643}{GW/m^3}$ at $y=\pm\SI{0.654}{mm}$ on quadrilaterals.
These values are very close, and the peaks are 0.026 mm apart.

For the steady flame, the integrated heat release rate is
$\SI{8.741}{kW/m}$ on triangles and $\SI{8.798}{kW/m}$ on
quadrilaterals, and the flame length is $\SI{6.050}{mm}$ and
$\SI{6.025}{mm}$, which differ by less than 1\% along the axis (Fig.~\ref{fig:burner-history}). For a slot
flame fed with a parabolic profile, Nambully et
al.~\cite{nambully2014} estimate the flame
length as:
\begin{equation}
  L_f=\frac{l_0}{\tan\beta_f},\qquad
  \sin\beta_f=\frac{S_L^0}{U_b},\qquad
  U_b=\frac{2u_0}{3},
  \label{eq:burner-lf}
\end{equation}
where $S_L^0$ is the laminar flame speed and $U_b$ the bulk velocity of
the inflow profile. With $S_L^0=\SI{0.2815}{m/s}$~\cite{marchal2023},
Eq.~\eqref{eq:burner-lf} gives $L_f=\SI{6.12}{mm}$, and both computed
lengths agree with this estimate (Fig.~\ref{fig:burner-history}b).

In this reacting test case, the NSCBC simplex formulation thus reproduces the
steady flame of the tensor-product formulation, with a similar transient
and a final flame whose length agrees with Eq.~\eqref{eq:burner-lf}.

%% file: figures/burner_sketch.tex
\begin{tikzpicture}[x=0.33cm,y=0.75cm,font=\footnotesize,
  wall/.style={line width=1.1pt},
  sym/.style={line width=0.6pt,dash pattern=on 4pt off 1.5pt on 1pt off 1.5pt},
  dim/.style={{Latex[length=1.6mm]}-{Latex[length=1.6mm]},line width=0.4pt}]
  \draw[wall] (0,0.65) -- (10,0.65) -- (10,1.7);
  \draw[wall] (0,-0.65) -- (10,-0.65) -- (10,-1.7);
  \draw[sym] (10,1.7) -- (30,1.7);
  \draw[sym] (10,-1.7) -- (30,-1.7);
  \draw[line width=0.8pt,blue!70!black] (0,-0.65) -- (0,0.65);
  \draw[line width=0.8pt,red!70!black] (30,-1.7) -- (30,1.7);
  \foreach \yy in {-0.5,-0.25,0,0.25,0.5} {
    \pgfmathsetmacro{\uu}{2.2*(1-(\yy/0.65)^2)}
    \draw[-{Latex[length=1.2mm]},blue!70!black,line width=0.4pt] (0,\yy) -- (\uu,\yy);
  }
  \draw[blue!70!black,line width=0.4pt,domain=-0.65:0.65,samples=30,smooth]
    plot ({2.2*(1-(\x/0.65)^2)},\x);
  \draw[orange!85!black,line width=1pt] (10,0.65) .. controls (13,0.55) and (15.6,0.25) .. (16.1,0)
    .. controls (15.6,-0.25) and (13,-0.55) .. (10,-0.65);
  \draw[gray,dashed,line width=0.5pt] (12,-1.7) -- (12,1.7);
  \node[gray,above] at (12,1.7) {$x=\SI{12}{mm}$};
  \node[anchor=east,blue!70!black,align=right] at (-0.3,0) {Inlet\\$u^\tgt(y)$};
  \node[anchor=west,red!70!black,align=left] at (30.3,0) {Outlet\\$P^\tgt$};
  \node[above] at (5,0.65) {No-slip wall};
  \node[above] at (21,1.7) {Symmetry plane};
  \node[below] at (21,-1.7) {Symmetry plane};
  \node[orange!85!black,anchor=south] at (15.0,0.45) {Flame};
  \draw[dim] (0,-2.35) -- node[below] {$\SI{10}{mm}$} (10,-2.35);
  \draw[dim] (10,-2.35) -- node[below] {$\SI{20}{mm}$} (30,-2.35);
  \draw[dim] (7,-0.65) -- node[left] {$\SI{1.3}{mm}$} (7,0.65);
  \draw[dim] (27,-1.7) -- node[left] {$\SI{3.4}{mm}$} (27,1.7);
  \draw[-{Latex[length=1.4mm]},line width=0.4pt] (19,-1.0) -- (21,-1.0) node[right] {$x$};
  \draw[-{Latex[length=1.4mm]},line width=0.4pt] (19,-1.0) -- (19,-0.1) node[above] {$y$};
\end{tikzpicture}

%% file: sections/07_conclusions.tex
\section{Conclusions and perspectives}
\label{sec:conclusions}

This work extends the NSCBC to the SD method on triangles and
tetrahedra. The proposed treatment performs the characteristic analysis in the physical frame of each NSCBC face FP, and evaluates the normal derivative of the normal flux with the RT divergence operator. The convective numerical flux at the NSCBC face FP then follows from a linear system local to each boundary element, which also imposes the viscous outflow conditions of Poinsot and
Lele~\cite{poinsot1992}. The matrix of this system depends on the
geometry only and can be inverted once during pre-processing. Moreover, the treatment applies unchanged to the SDRT and C-SDRT schemes. It also applies to multi-species reacting mixtures, for which the characteristic analysis only gains one wave per species, advected at the normal velocity (see~\ref{app:ms-nscbc}). To the authors' knowledge, this is the first NSCBC treatment for the SD method on simplex elements. Since it only requires the face frame, a divergence
operator evaluated at the FP and a local linear system, the same principle
should extend without major difficulty to the FR method on simplex
elements.

The treatment was validated against the tensor-product formulation of
Fiévet et al.~\cite{fievet2020} on six relevant test cases. An isentropic
vortex crossing a straight and an inclined outlet, a planar Poiseuille
flow and a two-dimensional premixed methane-air burner with multi-species chemistry were computed on triangles. In three dimensions, a spherical pressure wave in a cube and a Poiseuille flow in a pipe with curved walls were computed on tetrahedra. In all these considered cases, the results are very close to those obtained on quadrilaterals and
hexahedra.

The treatment of curved NSCBC faces remains to be assessed, since the
NSCBC faces of the test cases are planar and Eq.~\eqref{eq:inv-system} is
consistent with Eq.~\eqref{eq:bnd-coef} only when all the FP of the face
share the same normal. Elements with two or three NSCBC faces at the edges and corners of the domain are present in both the vortex and cube meshes and require no special treatment.
The wave strengths are built from the
Euler fluxes, following Poinsot and Lele, and including the diffusive
fluxes in them remains to be studied on simplex elements.
Future work will extend the treatment to the time-domain impedance boundary conditions that Fiévet et al.~\cite{fievet2020} derived on
tensor-product elements, to hybrid meshes in which NSCBC faces belong to
elements of different types, and to three-dimensional reacting flows on tetrahedra, and to industrial configurations.

%% file: sections/A_appendices.tex
\section{Transformation matrices in the face frame}
\label{app:matrices}

For a single-species calorically perfect gas, the matrices used in
Section~\ref{sec:simplex-char} read
\begin{equation}
  \dd{\tQ}{\tU} =
  \begin{pmatrix}
    1 & 0 & 0 & 0 & 0\\
    \displaystyle-\frac{u_n}{\rho} & \displaystyle\frac{1}{\rho} & 0 & 0 & 0\\
    \displaystyle-\frac{u_{t_1}}{\rho} & 0 & \displaystyle\frac{1}{\rho} & 0 & 0\\
    \displaystyle-\frac{u_{t_2}}{\rho} & 0 & 0 & \displaystyle\frac{1}{\rho} & 0\\
    \displaystyle\frac{(\gamma-1)\lVert\bu\rVert^2}{2} & (1-\gamma)u_n & (1-\gamma)u_{t_1} & (1-\gamma)u_{t_2} & \gamma-1
  \end{pmatrix},
  \label{eq:app-dQdU}
\end{equation}
\begin{equation}
  \dd{\tU}{\tQ} =
  \begin{pmatrix}
    1 & 0 & 0 & 0 & 0\\
    u_n & \rho & 0 & 0 & 0\\
    u_{t_1} & 0 & \rho & 0 & 0\\
    u_{t_2} & 0 & 0 & \rho & 0\\
    \displaystyle\frac{\lVert\bu\rVert^2}{2} & \rho u_n & \rho u_{t_1} & \rho u_{t_2} & \displaystyle\frac{1}{\gamma-1}
  \end{pmatrix},
  \label{eq:app-dUdQ}
\end{equation}
\begin{equation}
  \An =
  \begin{pmatrix}
    u_n & \rho & 0 & 0 & 0\\
    0 & u_n & 0 & 0 & 1/\rho\\
    0 & 0 & u_n & 0 & 0\\
    0 & 0 & 0 & u_n & 0\\
    0 & \gamma P & 0 & 0 & u_n
  \end{pmatrix},
  \label{eq:app-An}
\end{equation}
\begin{equation}
  \RAn =
  \begin{pmatrix}
    1 & 0 & 0 & \displaystyle\frac{\rho}{\sqrt{2}c} & \displaystyle\frac{\rho}{\sqrt{2}c}\\
    0 & 0 & 0 & \displaystyle\frac{1}{\sqrt{2}} & \displaystyle-\frac{1}{\sqrt{2}}\\
    0 & 0 & \displaystyle-\frac{c}{\rho} & 0 & 0\\
    0 & \displaystyle\frac{c}{\rho} & 0 & 0 & 0\\
    0 & 0 & 0 & \displaystyle\frac{\rho c}{\sqrt{2}} & \displaystyle\frac{\rho c}{\sqrt{2}}
  \end{pmatrix},
  \label{eq:app-R}
\end{equation}
\begin{equation}
  \LAn =
  \begin{pmatrix}
    1 & 0 & 0 & 0 & \displaystyle-\frac{1}{c^2}\\
    0 & 0 & 0 & \displaystyle\frac{\rho}{c} & 0\\
    0 & 0 & \displaystyle-\frac{\rho}{c} & 0 & 0\\
    0 & \displaystyle\frac{1}{\sqrt{2}} & 0 & 0 & \displaystyle\frac{1}{\sqrt{2}\rho c}\\
    0 & \displaystyle-\frac{1}{\sqrt{2}} & 0 & 0 & \displaystyle\frac{1}{\sqrt{2}\rho c}
  \end{pmatrix}.
  \label{eq:app-RL}
\end{equation}

\section{Tangential vectors}
\label{app:tangent}

For a unit normal $\bn=(n_x,n_y,n_z)^\T$, two unit tangential vectors are
built as follows. If $|n_z|<0.7$,
\begin{equation}
  \bt_1 = \frac{1}{\sqrt{n_x^2+n_y^2}}\begin{pmatrix} n_y\\ -n_x\\ 0\end{pmatrix},\qquad
  \bt_2 = \frac{1}{\sqrt{n_x^2+n_y^2}}\begin{pmatrix} n_x n_z\\ n_y n_z\\ -(n_x^2+n_y^2)\end{pmatrix},
\end{equation}
and otherwise
\begin{equation}
  \bt_1 = \frac{1}{\sqrt{n_y^2+n_z^2}}\begin{pmatrix} 0\\ -n_z\\ n_y\end{pmatrix},\qquad
  \bt_2 = \frac{1}{\sqrt{n_y^2+n_z^2}}\begin{pmatrix} n_y^2+n_z^2\\ -n_x n_y\\ -n_x n_z\end{pmatrix}.
\end{equation}
In both cases $\bt_2=\bn\times\bt_1$, so that $(\bn,\bt_1,\bt_2)$ is a
direct orthonormal frame.

\section{Extension to multi-species reacting flows}
\label{app:multispecies}

\subsection{Governing equations}
\label{app:ms-equations}

For a reacting mixture of $\Ns$ species, Eq.~\eqref{eq:nse-phys} becomes
\begin{equation}
  \dd{\bU}{t} + \divx\cF = \mathbf{S},
  \label{eq:ms-nse}
\end{equation}
with the conservative variables and the chemical source terms
\begin{equation}
  \bU = \big(\rho,\ \rho\bu^\T,\ \rho E,\ \rho Y_\spec\big)^\T,
  \qquad
  \mathbf{S} = \big(0,\ \mathbf{0}^\T,\ \dot{\omega}_T,\ \dot{\omega}_\spec\big)^\T,
  \label{eq:ms-US}
\end{equation}
for $\spec\in[1,\Ns]$, where $Y_\spec$ is the mass fraction of species
$\spec$ and $E$ the total non-chemical energy per unit mass,
\begin{equation}
  \rho E = \rho h_s - P + \frac{1}{2}\rho\lVert\bu\rVert^2,
  \label{eq:ms-energy}
\end{equation}
with the sensible enthalpies of the mixture and of species $\spec$,
\begin{equation}
  h_s = \sum_{\spec=1}^{\Ns} Y_\spec h_{s,\spec},
  \qquad
  h_{s,\spec}(T) = \int_{T_0}^{T} C_{p,\spec}(T')\,\mathrm{d}T',
  \label{eq:ms-hs}
\end{equation}
where $T_0$ is a reference temperature and $C_{p,\spec}$ the heat capacity
at constant pressure of species $\spec$.
The fluxes of Eq.~\eqref{eq:fluxes} receive one more row per species,
\begin{equation}
  \cF_c \ni \rho Y_\spec\,\bu^\T,
  \qquad
  \cF_d \ni \rho Y_\spec\big(\mathbf{V}_\spec+\mathbf{V}^c\big)^\T,
  \label{eq:ms-fluxes}
\end{equation}
and the heat flux becomes
\begin{equation}
  \mathbf{q} = -\lambda\gradx T
  + \rho\sum_{\spec=1}^{\Ns}h_{s,\spec}Y_\spec\big(\mathbf{V}_\spec+\mathbf{V}^c\big).
  \label{eq:ms-q}
\end{equation}
The diffusion velocities follow the Hirschfelder and Curtiss approximation
with a correction velocity that ensures mass conservation \cite{poinsot2005},
\begin{equation}
  \mathbf{V}_\spec = -\frac{D_\spec W_\spec}{Y_\spec W}\gradx X_\spec,
  \qquad
  \mathbf{V}^c = \sum_{\spec=1}^{\Ns} D_\spec\frac{W_\spec}{W}\gradx X_\spec,
  \label{eq:ms-diffusion}
\end{equation}
where $X_\spec$, $W_\spec$ and $D_\spec$ are the mole fraction, the molar
mass and the diffusion coefficient into the mixture of species $\spec$, and
$W$ the molar mass of the mixture, $1/W=\sum_\spec Y_\spec/W_\spec$. As in
\cite{marchal2023}, the viscosity follows the power law
$\mu=\mu_{\mathrm{ref}}(T/T_{\mathrm{ref}})^m$, with $\mu_{\mathrm{ref}}$ the
viscosity at the reference temperature $T_{\mathrm{ref}}$ and $m$ the
exponent. The thermal conductivity is $\lambda=\mu C_p/\mathrm{Pr}$ and the
diffusion coefficients are $D_\spec=\mu/(\rho\,\mathrm{Sc}_\spec)$, with a
constant Prandtl number $\mathrm{Pr}$ and constant Schmidt numbers
$\mathrm{Sc}_\spec$. The net production rate $\dot{\omega}_\spec$ of species
$\spec$ follows an Arrhenius law, and the heat release rate is
$\dot{\omega}_T=-\sum_\spec\Delta h^0_{f,\spec}\dot{\omega}_\spec$, with
$\Delta h^0_{f,\spec}$ the formation enthalpy of species $\spec$. The
mixture follows the ideal-gas law
\begin{equation}
  P = \rho\,\frac{\overline{R}}{W}\,T,
  \label{eq:ms-eos}
\end{equation}
with $\overline{R}$ the universal gas constant. As in \cite{marchal2023},
the density is not transported but recomputed as
$\rho=\sum_\spec\rho Y_\spec$.

\subsection{Numerical ingredients}
\label{app:ms-numerics}

The SD discretization of Section~\ref{sec:sd} extends to
Eq.~\eqref{eq:ms-nse} with one change, taken from Marchal et
al.~\cite{marchal2023}. The solution is extrapolated from the SP to the
FP in the primitive variables $(T,\bu,P,Y_\spec)$ instead of the
conservative variables. Marchal et al.\ call this approach TUPY and show
that, in a multi-species gas, extrapolating the conservative variables
creates pressure oscillations, because the pressure and the total energy
are not linearly related. Neither the positivity-preserving limiter for the mass fractions nor the SDLIFT formulation of the diffusive fluxes is used.

\subsection{Characteristic boundary conditions}
\label{app:ms-nscbc}

The face-frame vectors are extended with the species,
\begin{equation}
  \begin{gathered}
  \tU=\big(\rho,\ \rho u_n,\ \rho u_{t_1},\ \rho u_{t_2},\ \rho E,\ \rho Y_\spec\big)^\T,\\
  \tQ=\big(\rho,\ u_n,\ u_{t_1},\ u_{t_2},\ P,\ Y_\spec\big)^\T,
  \end{gathered}
\end{equation}
and the rotation matrix becomes
$\mathrm{diag}(\Mrot,I_{\Ns})$, with $\Mrot$ from Eq.~\eqref{eq:Mrot} and
$I_{\Ns}$ the identity of size $\Ns$. In the matrices below, the last row and the last column stand for one block per species. Consider $\Theta_k$ and $e_s$:
\begin{equation}
  \theta_\spec = h_{s,\spec}-TC_p\frac{W}{W_\spec},
  \qquad
  e_s = h_s-TC_p+\frac{\lVert\bu\rVert^2}{2},
  \label{eq:ms-theta}
\end{equation}
where $C_p$ is the heat capacity at constant pressure of the mixture and
$\gamma$ its heat capacity ratio. The passage matrices of
Eqs.~\eqref{eq:app-dQdU} and \eqref{eq:app-dUdQ} become
\begin{equation}
  \dd{\tQ}{\tU} =
  \begin{pmatrix}
    \Big(\dd{\tQ}{\tU}\Big)_{\!1} & (1-\gamma)\,\theta_\spec\,\mathbf{e}_5\\[6pt]
    -\dfrac{Y_\spec}{\rho}\,\mathbf{e}_1^\T & \dfrac{1}{\rho}
  \end{pmatrix},
  \label{eq:ms-dQdU}
\end{equation}
where $(\partial\tQ/\partial\tU)_1$ is the single-species matrix of
Eq.~\eqref{eq:app-dQdU}, unchanged, and $\mathbf{e}_1$, $\mathbf{e}_5$ are
the first and fifth unit vectors of $\mathbb{R}^5$. The inverse matrix reads
\begin{equation}
  \dd{\tU}{\tQ} =
  \begin{pmatrix}
    1 & 0 & 0 & 0 & 0 & 0\\
    u_n & \rho & 0 & 0 & 0 & 0\\
    u_{t_1} & 0 & \rho & 0 & 0 & 0\\
    u_{t_2} & 0 & 0 & \rho & 0 & 0\\
    e_s & \rho u_n & \rho u_{t_1} & \rho u_{t_2} & \frac{1}{\gamma-1} & \rho\,\theta_\spec\\
    Y_\spec & 0 & 0 & 0 & 0 & \rho
  \end{pmatrix}.
\end{equation}
The matrices $\An$, $\RAn$ and $\LAn$ of
Eqs.~\eqref{eq:app-An} to \eqref{eq:app-RL} are completed by a diagonal
block $u_n I_{\Ns}$ for $\An$ and $I_{\Ns}$ for $\RAn$ and $\LAn$, so that
\begin{equation}
  \LamAn = \mathrm{diag}\big(u_n,u_n,u_n,u_n+c,u_n-c,u_n I_{\Ns}\big).
\end{equation}
The species are thus advected at $u_n$ like the entropy and vorticity waves.
The primitive form Eq.~\eqref{eq:prim-explicit} gains one row per species,
\begin{equation}
  \dd{Y_\spec}{t} = -\big(\cL_{5+\spec}+\cT_{5+\spec}\big).
  \label{eq:ms-Y}
\end{equation}
At a subsonic inflow, $\cL^*_{5+\spec}$ are additional unknowns. Mass fractions are relaxed towards their targets $Y_\spec^\tgt$ with the coefficient
$K_Y$, which gives
\begin{equation}
  \cL^*_{5+\spec} = -\cT_{5+\spec}-K_Y\big(Y_\spec^\tgt-Y_\spec\big),
  \label{eq:ms-inY}
\end{equation}
and the temperature relation Eq.~\eqref{eq:in-L1} now accounts for the
variations of $W$ through Eq.~\eqref{eq:ms-eos},
\begin{multline}
  \cL^*_1 = -\cT_1+\frac{\rho}{P}\Bigg[\rho\Rgas K_T\big(T^\tgt-T\big)
    +\frac{\cL_4+\cL^*_5+\cT_4+\cT_5}{\sqrt{2}}\Big(\rho c-\frac{P}{c}\Big)\\
    -PW\sum_{\spec=1}^{\Ns}\frac{\cL^*_{5+\spec}+\cT_{5+\spec}}{W_\spec}\Bigg],
  \label{eq:ms-inL1}
\end{multline}
with $\Rgas=\overline{R}/W$.
The velocity relations of Eq.~\eqref{eq:in-un} and the outflow condition
Eq.~\eqref{eq:out} are unchanged. At a subsonic outflow, the viscous
conditions of Eq.~\eqref{eq:visc-out} are completed by one condition per
species on its diffusion flux \cite[Eq.~(7)]{sutherland2003},
\begin{equation}
  \dd{}{x_n}\Big(\rho Y_\spec\big(\mathbf{V}_\spec+\mathbf{V}^c\big)\cdot\bn\Big) = 0,
  \label{eq:ms-visc-out}
\end{equation}
imposed with the same linear system as in Section~\ref{sec:simplex-visc}.

\paragraph{Remark} The characteristic
analysis above is performed on the non-reacting equations: the source vector
$\mathbf{S}$ enters neither $\cL$ nor $\cT$. Baum et al. \cite{baum1995}
make the same assumption when they estimate the wave amplitudes of a
multicomponent reacting gas, while keeping the reaction terms in the
equations integrated in time. This assumption fails when a flame reaches the
boundary. Sutherland and Kennedy \cite{sutherland2003} show that a
non-reflecting condition neglecting the source terms then breaks down, and
that retaining them in the characteristic analysis avoids unphysical
pressure and velocity gradients and flow reversals. Yoo and Im
\cite{yoo2007} reach the same conclusion and include the reaction terms in
the incoming wave amplitudes. These terms are not included here, because no flame
approaches an NSCBC boundary in the cases of this paper: the chemical source
terms vanish in the boundary elements, and $\mathbf{S}$ plays no role in the
characteristic treatment. Extending the simplex formulation to flames
interacting with boundaries would require the corrections of
\cite{sutherland2003,yoo2007}.

%% file: main.bbl
\begin{thebibliography}{10}
\expandafter\ifx\csname url\endcsname\relax
  \def\url#1{\texttt{#1}}\fi
\expandafter\ifx\csname urlprefix\endcsname\relax\def\urlprefix{URL }\fi
\expandafter\ifx\csname href\endcsname\relax
  \def\href#1#2{#2} \def\path#1{#1}\fi

\bibitem{wang2013}
Z.~J. Wang, K.~Fidkowski, R.~Abgrall, F.~Bassi, D.~Caraeni, A.~Cary,
  H.~Deconinck, R.~Hartmann, K.~Hillewaert, H.~T. Huynh, et~al., High-order
  {CFD} methods: current status and perspective, International Journal for
  Numerical Methods in Fluids 72~(8) (2013) 811--845.
\newblock \href {https://doi.org/10.1002/fld.3767}
  {\path{doi:10.1002/fld.3767}}.

\bibitem{vanharen2017}
J.~Vanharen, G.~Puigt, X.~Vasseur, J.-F. Boussuge, P.~Sagaut, Revisiting the
  spectral analysis for high-order spectral discontinuous methods, Journal of
  Computational Physics 337 (2017) 379--402.
\newblock \href {https://doi.org/10.1016/j.jcp.2017.02.043}
  {\path{doi:10.1016/j.jcp.2017.02.043}}.

\bibitem{cockburn2001}
B.~Cockburn, C.-W. Shu, {Runge-Kutta} discontinuous {Galerkin} methods for
  convection-dominated problems, Journal of Scientific Computing 16~(3) (2001)
  173--261.
\newblock \href {https://doi.org/10.1023/A:1012873910884}
  {\path{doi:10.1023/A:1012873910884}}.

\bibitem{bassi1997}
F.~Bassi, S.~Rebay, A high-order accurate discontinuous finite element method
  for the numerical solution of the compressible {Navier-Stokes} equations,
  Journal of Computational Physics 131~(2) (1997) 267--279.
\newblock \href {https://doi.org/10.1006/jcph.1996.5572}
  {\path{doi:10.1006/jcph.1996.5572}}.

\bibitem{hesthaven2008}
J.~S. Hesthaven, T.~Warburton, Nodal Discontinuous {Galerkin} Methods:
  Algorithms, Analysis, and Applications, Texts in Applied Mathematics,
  Springer, New York, 2008.
\newblock \href {https://doi.org/10.1007/978-0-387-72067-8}
  {\path{doi:10.1007/978-0-387-72067-8}}.

\bibitem{yu2014}
M.~Yu, Z.~J. Wang, Y.~Liu, On the accuracy and efficiency of discontinuous
  {Galerkin}, spectral difference and correction procedure via reconstruction
  methods, Journal of Computational Physics 259 (2014) 70--95.
\newblock \href {https://doi.org/10.1016/j.jcp.2013.11.023}
  {\path{doi:10.1016/j.jcp.2013.11.023}}.

\bibitem{kopriva2009}
D.~A. Kopriva, Implementing Spectral Methods for Partial Differential
  Equations: Algorithms for Scientists and Engineers, Scientific Computation,
  Springer Netherlands, 2009.
\newblock \href {https://doi.org/10.1007/978-90-481-2261-5}
  {\path{doi:10.1007/978-90-481-2261-5}}.

\bibitem{gassner2013}
G.~J. Gassner, A skew-symmetric discontinuous {Galerkin} spectral element
  discretization and its relation to {SBP-SAT} finite difference methods, SIAM
  Journal on Scientific Computing 35~(3) (2013) A1233--A1253.
\newblock \href {https://doi.org/10.1137/120890144}
  {\path{doi:10.1137/120890144}}.

\bibitem{carpenter2014}
M.~H. Carpenter, T.~C. Fisher, E.~J. Nielsen, S.~H. Frankel, Entropy stable
  spectral collocation schemes for the {Navier-Stokes} equations: discontinuous
  interfaces, SIAM Journal on Scientific Computing 36~(5) (2014) B835--B867.
\newblock \href {https://doi.org/10.1137/130932193}
  {\path{doi:10.1137/130932193}}.

\bibitem{gassner2016}
G.~J. Gassner, A.~R. Winters, D.~A. Kopriva, Split form nodal discontinuous
  {Galerkin} schemes with summation-by-parts property for the compressible
  {Euler} equations, Journal of Computational Physics 327 (2016) 39--66.
\newblock \href {https://doi.org/10.1016/j.jcp.2016.09.013}
  {\path{doi:10.1016/j.jcp.2016.09.013}}.

\bibitem{schwarz2025}
A.~Schwarz, D.~Kempf, J.~Keim, P.~Kopper, C.~Rohde, A.~Beck, Comparison of
  entropy stable collocation high-order {DG} methods for compressible turbulent
  flows, Computers \& Fluids 303 (2025) 106874.
\newblock \href {https://doi.org/10.1016/j.compfluid.2025.106874}
  {\path{doi:10.1016/j.compfluid.2025.106874}}.

\bibitem{kopriva1996}
D.~A. Kopriva, J.~H. Kolias, A conservative staggered-grid {Chebyshev}
  multidomain method for compressible flows, Journal of Computational Physics
  125~(1) (1996) 244--261.
\newblock \href {https://doi.org/10.1006/jcph.1996.0091}
  {\path{doi:10.1006/jcph.1996.0091}}.

\bibitem{kopriva1998}
D.~A. Kopriva, A staggered-grid multidomain spectral method for the
  compressible {Navier-Stokes} equations, Journal of Computational Physics
  143~(1) (1998) 125--158.
\newblock \href {https://doi.org/10.1006/jcph.1998.5956}
  {\path{doi:10.1006/jcph.1998.5956}}.

\bibitem{liu2006}
Y.~Liu, M.~Vinokur, Z.~J. Wang, Spectral difference method for unstructured
  grids {I}: basic formulation, Journal of Computational Physics 216~(2) (2006)
  780--801.
\newblock \href {https://doi.org/10.1016/j.jcp.2006.01.024}
  {\path{doi:10.1016/j.jcp.2006.01.024}}.

\bibitem{may2006}
G.~May, A.~Jameson, A spectral difference method for the {Euler} and
  {Navier-Stokes} equations on unstructured meshes, in: 44th {AIAA} {Aerospace
  Sciences Meeting and Exhibit}, 2006, {AIAA} Paper 2006-304.
\newblock \href {https://doi.org/10.2514/6.2006-304}
  {\path{doi:10.2514/6.2006-304}}.

\bibitem{sun2007}
Y.~Sun, Z.~J. Wang, Y.~Liu, High-order multidomain spectral difference method
  for the {Navier-Stokes} equations on unstructured hexahedral grids,
  Communications in Computational Physics 2~(2) (2007) 310--333.
\newblock \href {https://doi.org/10.4208/cicp.2007.v2.p310}
  {\path{doi:10.4208/cicp.2007.v2.p310}}.

\bibitem{huynh2007}
H.~T. Huynh, A flux reconstruction approach to high-order schemes including
  discontinuous {Galerkin} methods, in: 18th {AIAA} {Computational Fluid
  Dynamics Conference}, 2007, {AIAA} Paper 2007-4079.
\newblock \href {https://doi.org/10.2514/6.2007-4079}
  {\path{doi:10.2514/6.2007-4079}}.

\bibitem{wang2009}
Z.~J. Wang, H.~Gao, A unifying lifting collocation penalty formulation
  including the discontinuous {Galerkin}, spectral volume/difference methods
  for conservation laws on mixed grids, Journal of Computational Physics
  228~(21) (2009) 8161--8186.
\newblock \href {https://doi.org/10.1016/j.jcp.2009.07.036}
  {\path{doi:10.1016/j.jcp.2009.07.036}}.

\bibitem{vincent2011}
P.~E. Vincent, P.~Castonguay, A.~Jameson, A new class of high-order energy
  stable flux reconstruction schemes, Journal of Scientific Computing 47~(1)
  (2011) 50--72.
\newblock \href {https://doi.org/10.1007/s10915-010-9420-z}
  {\path{doi:10.1007/s10915-010-9420-z}}.

\bibitem{cox2021}
C.~Cox, W.~Trojak, T.~Dzanic, F.~D. Witherden, A.~Jameson, Accuracy, stability,
  and performance comparison between the spectral difference and flux
  reconstruction schemes, Computers \& Fluids 221 (2021) 104922.
\newblock \href {https://doi.org/10.1016/j.compfluid.2021.104922}
  {\path{doi:10.1016/j.compfluid.2021.104922}}.

\bibitem{huynh2009}
H.~T. Huynh, A reconstruction approach to high-order schemes including
  discontinuous {Galerkin} for diffusion, in: 47th {AIAA} {Aerospace Sciences
  Meeting including the New Horizons Forum and Aerospace Exposition}, 2009,
  {AIAA} Paper 2009-403.
\newblock \href {https://doi.org/10.2514/6.2009-403}
  {\path{doi:10.2514/6.2009-403}}.

\bibitem{lodato2014}
G.~Lodato, P.~Castonguay, A.~Jameson, Structural wall-modeled {LES} using a
  high-order spectral difference scheme for unstructured meshes, Flow,
  Turbulence and Combustion 92~(1-2) (2014) 579--606.
\newblock \href {https://doi.org/10.1007/s10494-013-9523-3}
  {\path{doi:10.1007/s10494-013-9523-3}}.

\bibitem{chapelier2016}
J.-B. Chapelier, G.~Lodato, A spectral-element dynamic model for the large-eddy
  simulation of turbulent flows, Journal of Computational Physics 321 (2016)
  279--302.
\newblock \href {https://doi.org/10.1016/j.jcp.2016.05.051}
  {\path{doi:10.1016/j.jcp.2016.05.051}}.

\bibitem{pinto2019}
B.~Pinto, G.~Lodato, Synthetic freestream disturbance for the numerical
  reproduction of experimental zero-pressure-gradient bypass transition test
  cases, Flow, Turbulence and Combustion 103~(1) (2019) 25--54.
\newblock \href {https://doi.org/10.1007/s10494-018-0004-6}
  {\path{doi:10.1007/s10494-018-0004-6}}.

\bibitem{lodato2021}
G.~Lodato, N.~Tonicello, B.~Pinto, Large-eddy simulation of bypass transition
  on a zero-pressure-gradient flat plate using the spectral-element dynamic
  model, Flow, Turbulence and Combustion 107~(4) (2021) 845--874.
\newblock \href {https://doi.org/10.1007/s10494-021-00262-1}
  {\path{doi:10.1007/s10494-021-00262-1}}.

\bibitem{gao2017}
J.~Gao, X.~Li, D.~Lin, Numerical simulation of the noise from the {30P30N}
  high-lift airfoil with spectral difference method, in: 23rd {AIAA/CEAS
  Aeroacoustics Conference}, 2017, {AIAA} Paper 2017-3363.
\newblock \href {https://doi.org/10.2514/6.2017-3363}
  {\path{doi:10.2514/6.2017-3363}}.

\bibitem{fievet2020hyp}
R.~Fiévet, H.~Deniau, J.-P. Brazier, E.~Piot, Numerical study of hypersonic
  boundary-layer transition delay through second-mode absorption, in: {AIAA
  SciTech} 2020 Forum, 2020, {AIAA} Paper 2020-2061.
\newblock \href {https://doi.org/10.2514/6.2020-2061}
  {\path{doi:10.2514/6.2020-2061}}.

\bibitem{lodato2019}
G.~Lodato, Characteristic modal shock detection for discontinuous finite
  element methods, Computers \& Fluids 179 (2019) 309--333.
\newblock \href {https://doi.org/10.1016/j.compfluid.2018.11.008}
  {\path{doi:10.1016/j.compfluid.2018.11.008}}.

\bibitem{tonicello2020}
N.~Tonicello, G.~Lodato, L.~Vervisch, Entropy preserving low dissipative shock
  capturing with wave-characteristic based sensor for high-order methods,
  Computers \& Fluids 197 (2020) 104357.
\newblock \href {https://doi.org/10.1016/j.compfluid.2019.104357}
  {\path{doi:10.1016/j.compfluid.2019.104357}}.

\bibitem{messai2024sc}
N.-A. Messaï, G.~Daviller, J.-F. Boussuge, Artificial viscosity-based shock
  capturing scheme for the spectral difference method on simplicial elements,
  Journal of Computational Physics 504 (2024) 112864.
\newblock \href {https://doi.org/10.1016/j.jcp.2024.112864}
  {\path{doi:10.1016/j.jcp.2024.112864}}.

\bibitem{marchal2025}
T.~Marchal, H.~Deniau, G.~Puigt, Entropy-stable spectral difference and flux
  reconstruction methods for discontinuous flows, Journal of Computational
  Physics 527 (2025) 113803.
\newblock \href {https://doi.org/10.1016/j.jcp.2025.113803}
  {\path{doi:10.1016/j.jcp.2025.113803}}.

\bibitem{chapelier2024}
J.-B. Chapelier, D.~J. Lusher, W.~Van~Noordt, C.~Wenzel, T.~Gibis, P.~Mossier,
  A.~Beck, G.~Lodato, C.~Brehm, M.~Ruggeri, C.~Scalo, N.~Sandham, Comparison of
  high-order numerical methodologies for the simulation of the supersonic
  {Taylor-Green} vortex flow, Physics of Fluids 36~(5) (2024) 055146.
\newblock \href {https://doi.org/10.1063/5.0206359}
  {\path{doi:10.1063/5.0206359}}.

\bibitem{gao2026}
J.~Gao, Z.~Jia, J.~Zhao, Numerical simulation of supersonic jet noise from a
  chevron nozzle with spectral difference method, in: 32nd {AIAA/CEAS
  Aeroacoustics Conference}, 2026, {AIAA Paper} 2026-3441.
\newblock \href {https://doi.org/10.2514/6.2026-3441}
  {\path{doi:10.2514/6.2026-3441}}.

\bibitem{brunet2018}
V.~Brunet, E.~Croner, A.~Minot, J.~De~Laborderie, E.~Lippinois, S.~Richard,
  J.-F. Boussuge, J.~Dombard, F.~Duchaine, L.~Gicquel, T.~Poinsot, G.~Puigt,
  G.~Staffelbach, L.~Segui, O.~Vermorel, N.~Villedieu, J.-S. Cagnone,
  K.~Hillewaert, M.~Rasquin, G.~Lartigue, V.~Moureau, V.~Couaillier, E.~Martin,
  M.~De~La Llave~Plata, J.-M. Le~Gouez, F.~Renac, Comparison of various {CFD}
  codes for {LES} simulations of turbomachinery: from inviscid vortex
  convection to multi-stage compressor, in: {ASME Turbo Expo} 2018, Volume
  {2C}: Turbomachinery, 2018, p. V02CT42A013.
\newblock \href {https://doi.org/10.1115/GT2018-75523}
  {\path{doi:10.1115/GT2018-75523}}.

\bibitem{tonicello2026}
N.~Tonicello, G.~Lodato, M.~Ihme, Extension of a spectral difference method for
  the diffused-interface five-equation model, Computers \& Fluids 304 (2026)
  106880.
\newblock \href {https://doi.org/10.1016/j.compfluid.2025.106880}
  {\path{doi:10.1016/j.compfluid.2025.106880}}.

\bibitem{gupta2018}
P.~Gupta, K.~Schwinn, G.~Lodato, C.~Slabaugh, C.~Scalo, Numerical investigation
  of sustained planar detonation waves in a periodic domain, in: 2018 Fluid
  Dynamics Conference, 2018, {AIAA} Paper 2018-3240.
\newblock \href {https://doi.org/10.2514/6.2018-3240}
  {\path{doi:10.2514/6.2018-3240}}.

\bibitem{tofaili2021}
H.~Tofaili, G.~Lodato, L.~Vervisch, P.~Clavin, One-dimensional dynamics of
  gaseous detonations revisited, Combustion and Flame 232 (2021) 111535.
\newblock \href {https://doi.org/10.1016/j.combustflame.2021.111535}
  {\path{doi:10.1016/j.combustflame.2021.111535}}.

\bibitem{marchal2023}
T.~Marchal, H.~Deniau, J.-F. Boussuge, B.~Cuenot, R.~Mercier, Extension of the
  spectral difference method to premixed laminar and turbulent combustion,
  Flow, Turbulence and Combustion 111~(1) (2023) 141--176.
\newblock \href {https://doi.org/10.1007/s10494-023-00414-5}
  {\path{doi:10.1007/s10494-023-00414-5}}.

\bibitem{castro2026}
T.~Castro, T.~Marchal, Simulations of rim stabilized edge flames at atmospheric
  and high pressures using a p-adaptive spectral difference method, Physics of
  Fluids 38~(1) (2026) 016105.
\newblock \href {https://doi.org/10.1063/5.0303121}
  {\path{doi:10.1063/5.0303121}}.

\bibitem{lu2015}
Y.~Lu, W.~N. Dawes, High order large eddy simulations for a transonic turbine
  blade using hybrid unstructured meshes, in: {ASME Turbo Expo} 2015, Volume
  {2C}: Turbomachinery, 2015, p. V02CT44A005.
\newblock \href {https://doi.org/10.1115/GT2015-42283}
  {\path{doi:10.1115/GT2015-42283}}.

\bibitem{priebe2022}
S.~Priebe, T.~Wood, J.~Yi, A.~Mousavi, Large eddy simulation of an open rotor
  fan blade, in: {ASME Turbo Expo} 2022, Volume {10A}: Turbomachinery, 2022, p.
  V10AT29A009.
\newblock \href {https://doi.org/10.1115/GT2022-80538}
  {\path{doi:10.1115/GT2022-80538}}.

\bibitem{priebe2025}
S.~Priebe, E.~Jourdan, A.~Mousavi, R.~Karve, S.~Chakrabarti, L.~D'Aquila,
  J.~Yi, M.~Alhawwary, V.~Bharadwaj~Ananthan, K.~Ramakrishnan, T.~Wood, Large
  eddy simulation of an open fan blade at full-scale {Reynolds} number, Journal
  of Turbomachinery 147~(7) (2025) 071004.
\newblock \href {https://doi.org/10.1115/1.4067098}
  {\path{doi:10.1115/1.4067098}}.

\bibitem{wang2022}
Z.~J. Wang, High order wall-modeled large-eddy simulation on mixed unstructured
  meshes, {AIAA} Journal 60~(12) (2022) 6881--6896.
\newblock \href {https://doi.org/10.2514/1.J061641}
  {\path{doi:10.2514/1.J061641}}.

\bibitem{wang2024}
Z.~J. Wang, High-order wall-modeled large-eddy simulation of high-lift
  configuration, {AIAA} Journal 62~(9) (2024) 3326--3339.
\newblock \href {https://doi.org/10.2514/1.J063332}
  {\path{doi:10.2514/1.J063332}}.

\bibitem{ching2019}
E.~J. Ching, Y.~Lv, P.~Gnoffo, M.~Barnhardt, M.~Ihme, Shock capturing for
  discontinuous {Galerkin} methods with application to predicting heat transfer
  in hypersonic flows, Journal of Computational Physics 376 (2019) 54--75.
\newblock \href {https://doi.org/10.1016/j.jcp.2018.09.016}
  {\path{doi:10.1016/j.jcp.2018.09.016}}.

\bibitem{hoskin2024}
D.~S. Hoskin, R.~L. Van~Heyningen, N.~C. Nguyen, J.~Vila-Pérez, W.~L. Harris,
  J.~Peraire, Discontinuous {Galerkin} methods for hypersonic flows, Progress
  in Aerospace Sciences 146 (2024) 100999.
\newblock \href {https://doi.org/10.1016/j.paerosci.2024.100999}
  {\path{doi:10.1016/j.paerosci.2024.100999}}.

\bibitem{garai2016}
A.~Garai, L.~T. Diosady, S.~M. Murman, N.~K. Madavan, {DNS} of low-pressure
  turbine cascade flows with elevated inflow turbulence using a
  discontinuous-{Galerkin} spectral-element method, in: ASME Turbo Expo 2016,
  Volume 2C: Turbomachinery, 2016, p. V02CT39A025.
\newblock \href {https://doi.org/10.1115/GT2016-56700}
  {\path{doi:10.1115/GT2016-56700}}.

\bibitem{lv2014}
Y.~Lv, M.~Ihme, Discontinuous {Galerkin} method for multicomponent chemically
  reacting flows and combustion, Journal of Computational Physics 270 (2014)
  105--137.
\newblock \href {https://doi.org/10.1016/j.jcp.2014.03.029}
  {\path{doi:10.1016/j.jcp.2014.03.029}}.

\bibitem{billet2014}
G.~Billet, J.~Ryan, M.~Borrel, A {Runge-Kutta} discontinuous {Galerkin}
  approach to solve reactive flows on conforming hybrid grids: the parabolic
  and source operators, Computers \& Fluids 95 (2014) 98--115.
\newblock \href {https://doi.org/10.1016/j.compfluid.2014.01.031}
  {\path{doi:10.1016/j.compfluid.2014.01.031}}.

\bibitem{johnson2020}
R.~F. Johnson, A.~D. Kercher, A conservative discontinuous {Galerkin}
  discretization for the chemically reacting {Navier-Stokes} equations, Journal
  of Computational Physics 423 (2020) 109826.
\newblock \href {https://doi.org/10.1016/j.jcp.2020.109826}
  {\path{doi:10.1016/j.jcp.2020.109826}}.

\bibitem{castonguay2012}
P.~Castonguay, P.~E. Vincent, A.~Jameson, A new class of high-order energy
  stable flux reconstruction schemes for triangular elements, Journal of
  Scientific Computing 51~(1) (2012) 224--256.
\newblock \href {https://doi.org/10.1007/s10915-011-9505-3}
  {\path{doi:10.1007/s10915-011-9505-3}}.

\bibitem{williams2014fr}
D.~M. Williams, A.~Jameson, Energy stable flux reconstruction schemes for
  advection-diffusion problems on tetrahedra, Journal of Scientific Computing
  59~(3) (2014) 721--759.
\newblock \href {https://doi.org/10.1007/s10915-013-9780-2}
  {\path{doi:10.1007/s10915-013-9780-2}}.

\bibitem{balan2012}
A.~Balan, G.~May, J.~Sch\"oberl, A stable high-order spectral difference method
  for hyperbolic conservation laws on triangular elements, Journal of
  Computational Physics 231~(5) (2012) 2359--2375.
\newblock \href {https://doi.org/10.1016/j.jcp.2011.11.041}
  {\path{doi:10.1016/j.jcp.2011.11.041}}.

\bibitem{veilleux2022jcp}
A.~Veilleux, G.~Puigt, H.~Deniau, G.~Daviller, A stable {Spectral Difference}
  approach for computations with triangular and hybrid grids up to the 6th
  order of accuracy, Journal of Computational Physics 449 (2022) 110774.
\newblock \href {https://doi.org/10.1016/j.jcp.2021.110774}
  {\path{doi:10.1016/j.jcp.2021.110774}}.

\bibitem{veilleux2022jsc}
A.~Veilleux, G.~Puigt, H.~Deniau, G.~Daviller, Stable {Spectral Difference}
  approach using {Raviart-Thomas} elements for {3D} computations on tetrahedral
  grids, Journal of Scientific Computing 91 (2022) 7.
\newblock \href {https://doi.org/10.1007/s10915-022-01790-2}
  {\path{doi:10.1007/s10915-022-01790-2}}.

\bibitem{messai2024}
N.-A. Messa\"i, G.~Daviller, A corrected {Raviart-Thomas} {Spectral Difference}
  scheme stable for arbitrary order of accuracy on triangular and tetrahedral
  meshes, Computer Methods in Applied Mechanics and Engineering 432 (2024)
  117413.
\newblock \href {https://doi.org/10.1016/j.cma.2024.117413}
  {\path{doi:10.1016/j.cma.2024.117413}}.

\bibitem{poinsot1992}
T.~J. Poinsot, S.~K. Lele, Boundary conditions for direct simulations of
  compressible viscous flows, Journal of Computational Physics 101~(1) (1992)
  104--129.
\newblock \href {https://doi.org/10.1016/0021-9991(92)90046-2}
  {\path{doi:10.1016/0021-9991(92)90046-2}}.

\bibitem{thompson1987}
K.~W. Thompson, Time dependent boundary conditions for hyperbolic systems,
  Journal of Computational Physics 68~(1) (1987) 1--24.
\newblock \href {https://doi.org/10.1016/0021-9991(87)90041-6}
  {\path{doi:10.1016/0021-9991(87)90041-6}}.

\bibitem{rudy1980}
D.~H. Rudy, J.~C. Strikwerda, A nonreflecting outflow boundary condition for
  subsonic {Navier-Stokes} calculations, Journal of Computational Physics
  36~(1) (1980) 55--70.
\newblock \href {https://doi.org/10.1016/0021-9991(80)90174-6}
  {\path{doi:10.1016/0021-9991(80)90174-6}}.

\bibitem{selle2004}
L.~Selle, F.~Nicoud, T.~Poinsot, Actual impedance of nonreflecting boundary
  conditions: implications for computation of resonators, {AIAA} Journal 42~(5)
  (2004) 958--964.
\newblock \href {https://doi.org/10.2514/1.1883} {\path{doi:10.2514/1.1883}}.

\bibitem{polifke2006}
W.~Polifke, C.~Wall, P.~Moin, Partially reflecting and non-reflecting boundary
  conditions for simulation of compressible viscous flow, Journal of
  Computational Physics 213~(1) (2006) 437--449.
\newblock \href {https://doi.org/10.1016/j.jcp.2005.08.016}
  {\path{doi:10.1016/j.jcp.2005.08.016}}.

\bibitem{baum1995}
M.~Baum, T.~Poinsot, D.~Th\'evenin, Accurate boundary conditions for
  multicomponent reactive flows, Journal of Computational Physics 116~(2)
  (1995) 247--261.
\newblock \href {https://doi.org/10.1006/jcph.1995.1024}
  {\path{doi:10.1006/jcph.1995.1024}}.

\bibitem{sutherland2003}
J.~C. Sutherland, C.~A. Kennedy, Improved boundary conditions for viscous,
  reacting, compressible flows, Journal of Computational Physics 191~(2) (2003)
  502--524.
\newblock \href {https://doi.org/10.1016/S0021-9991(03)00328-0}
  {\path{doi:10.1016/S0021-9991(03)00328-0}}.

\bibitem{yoo2005}
C.~S. Yoo, Y.~Wang, A.~Trouv\'e, H.~G. Im, Characteristic boundary conditions
  for direct simulations of turbulent counterflow flames, Combustion Theory and
  Modelling 9~(4) (2005) 617--646.
\newblock \href {https://doi.org/10.1080/13647830500307378}
  {\path{doi:10.1080/13647830500307378}}.

\bibitem{yoo2007}
C.~S. Yoo, H.~G. Im, Characteristic boundary conditions for simulations of
  compressible reacting flows with multi-dimensional, viscous and reaction
  effects, Combustion Theory and Modelling 11~(2) (2007) 259--286.
\newblock \href {https://doi.org/10.1080/13647830600898995}
  {\path{doi:10.1080/13647830600898995}}.

\bibitem{lodato2008}
G.~Lodato, P.~Domingo, L.~Vervisch, Three-dimensional boundary conditions for
  direct and large-eddy simulation of compressible viscous flows, Journal of
  Computational Physics 227~(10) (2008) 5105--5143.
\newblock \href {https://doi.org/10.1016/j.jcp.2008.01.038}
  {\path{doi:10.1016/j.jcp.2008.01.038}}.

\bibitem{kim2000}
J.~W. Kim, D.~J. Lee, Generalized characteristic boundary conditions for
  computational aeroacoustics, {AIAA} Journal 38~(11) (2000) 2040--2049.
\newblock \href {https://doi.org/10.2514/2.891} {\path{doi:10.2514/2.891}}.

\bibitem{kim2004}
J.~W. Kim, D.~J. Lee, Generalized characteristic boundary conditions for
  computational aeroacoustics, {Part} 2, {AIAA} Journal 42~(1) (2004) 47--55.
\newblock \href {https://doi.org/10.2514/1.9029} {\path{doi:10.2514/1.9029}}.

\bibitem{granet2010}
V.~Granet, O.~Vermorel, T.~L\'eonard, L.~Gicquel, T.~Poinsot, Comparison of
  nonreflecting outlet boundary conditions for compressible solvers on
  unstructured grids, {AIAA} Journal 48~(10) (2010) 2348--2364.
\newblock \href {https://doi.org/10.2514/1.J050391}
  {\path{doi:10.2514/1.J050391}}.

\bibitem{daviller2019}
G.~Daviller, G.~Oztarlik, T.~Poinsot, A generalized non-reflecting inlet
  boundary condition for steady and forced compressible flows with injection of
  vortical and acoustic waves, Computers \& Fluids 190 (2019) 503--513.
\newblock \href {https://doi.org/10.1016/j.compfluid.2019.06.027}
  {\path{doi:10.1016/j.compfluid.2019.06.027}}.

\bibitem{fosso2012}
A.~Fosso~Pouangu\'e, H.~Deniau, N.~Lamarque, T.~Poinsot, Comparison of outflow
  boundary conditions for subsonic aeroacoustic simulations, International
  Journal for Numerical Methods in Fluids 68~(10) (2012) 1207--1233.
\newblock \href {https://doi.org/10.1002/fld.2597}
  {\path{doi:10.1002/fld.2597}}.

\bibitem{fievet2020}
R.~Fi\'evet, H.~Deniau, E.~Piot, Strong compact formalism for characteristic
  boundary conditions with discontinuous spectral methods, Journal of
  Computational Physics 408 (2020) 109276.
\newblock \href {https://doi.org/10.1016/j.jcp.2020.109276}
  {\path{doi:10.1016/j.jcp.2020.109276}}.

\bibitem{toulopoulos2011}
I.~Toulopoulos, J.~A. Ekaterinaris, Artificial boundary conditions for the
  numerical solution of the {Euler} equations by the discontinuous {Galerkin}
  method, Journal of Computational Physics 230~(15) (2011) 5974--5995.
\newblock \href {https://doi.org/10.1016/j.jcp.2011.04.008}
  {\path{doi:10.1016/j.jcp.2011.04.008}}.

\bibitem{shehadi2024}
E.~K. Shehadi, E.~T.~A. {van der Weide}, A polynomial-correction
  {Navier-Stokes} characteristic boundary condition, Computer Physics
  Communications 303 (2024) 109194.
\newblock \href {https://doi.org/10.1016/j.cpc.2024.109194}
  {\path{doi:10.1016/j.cpc.2024.109194}}.

\bibitem{ellmenreich2026}
J.~Ellmenreich, M.~Giacomini, A.~Huerta, P.~L. Lederer, Characteristic boundary
  conditions for hybridizable discontinuous {Galerkin} methods, Journal of
  Computational Physics 548 (2026) 114565.
\newblock \href {https://doi.org/10.1016/j.jcp.2025.114565}
  {\path{doi:10.1016/j.jcp.2025.114565}}.

\bibitem{kopriva2006}
D.~A. Kopriva, Metric identities and the discontinuous spectral element method
  on curvilinear meshes, Journal of Scientific Computing 26~(3) (2006)
  301--327.
\newblock \href {https://doi.org/10.1007/s10915-005-9070-8}
  {\path{doi:10.1007/s10915-005-9070-8}}.

\bibitem{abe2015}
Y.~Abe, T.~Haga, T.~Nonomura, K.~Fujii, On the freestream preservation of
  high-order conservative flux-reconstruction schemes, Journal of Computational
  Physics 281 (2015) 28--54.
\newblock \href {https://doi.org/10.1016/j.jcp.2014.10.011}
  {\path{doi:10.1016/j.jcp.2014.10.011}}.

\bibitem{vandenabeele2008}
K.~Van~den Abeele, C.~Lacor, Z.~J. Wang, On the stability and accuracy of the
  spectral difference method, Journal of Scientific Computing 37~(2) (2008)
  162--188.
\newblock \href {https://doi.org/10.1007/s10915-008-9201-0}
  {\path{doi:10.1007/s10915-008-9201-0}}.

\bibitem{jameson2010}
A.~Jameson, A proof of the stability of the spectral difference method for all
  orders of accuracy, Journal of Scientific Computing 45~(1-3) (2010) 348--358.
\newblock \href {https://doi.org/10.1007/s10915-009-9339-4}
  {\path{doi:10.1007/s10915-009-9339-4}}.

\bibitem{williams2014}
D.~M. Williams, L.~Shunn, A.~Jameson, Symmetric quadrature rules for simplexes
  based on sphere close packed lattice arrangements, Journal of Computational
  and Applied Mathematics 266 (2014) 18--38.
\newblock \href {https://doi.org/10.1016/j.cam.2014.01.007}
  {\path{doi:10.1016/j.cam.2014.01.007}}.

\bibitem{shunn2012}
L.~Shunn, F.~Ham, Symmetric quadrature rules for tetrahedra based on a cubic
  close-packed lattice arrangement, Journal of Computational and Applied
  Mathematics 236~(17) (2012) 4348--4364.
\newblock \href {https://doi.org/10.1016/j.cam.2012.03.032}
  {\path{doi:10.1016/j.cam.2012.03.032}}.

\bibitem{koornwinder1975}
T.~Koornwinder, Two-variable analogues of the classical orthogonal polynomials,
  in: R.~A. Askey (Ed.), Theory and Application of Special Functions, Academic
  Press, 1975, pp. 435--495.
\newblock \href {https://doi.org/10.1016/B978-0-12-064850-4.50015-X}
  {\path{doi:10.1016/B978-0-12-064850-4.50015-X}}.

\bibitem{dubiner1991}
M.~Dubiner, Spectral methods on triangles and other domains, Journal of
  Scientific Computing 6~(4) (1991) 345--390.
\newblock \href {https://doi.org/10.1007/BF01060030}
  {\path{doi:10.1007/BF01060030}}.

\bibitem{batten1997}
P.~Batten, N.~Clarke, C.~Lambert, D.~M. Causon, On the choice of wavespeeds for
  the {HLLC} {R}iemann solver, SIAM Journal on Scientific Computing 18~(6)
  (1997) 1553--1570.
\newblock \href {https://doi.org/10.1137/S1064827593260140}
  {\path{doi:10.1137/S1064827593260140}}.

\bibitem{gottlieb1998}
S.~Gottlieb, C.-W. Shu, Total variation diminishing {R}unge-{K}utta schemes,
  Mathematics of Computation 67~(221) (1998) 73--85.
\newblock \href {https://doi.org/10.1090/S0025-5718-98-00913-2}
  {\path{doi:10.1090/S0025-5718-98-00913-2}}.

\bibitem{Bogey2002e}
C.~Bogey, C.~Bailly, {Three-dimensional non-reflective boundary conditions for
  acoustic simulations: far field formulation and validation test cases}, Acta
  Acustica united with Acustica 88 (2002) 463--471.

\bibitem{spiteri2002}
R.~J. Spiteri, S.~J. Ruuth, A new class of optimal high-order
  strong-stability-preserving time discretization methods, SIAM Journal on
  Numerical Analysis 40~(2) (2002) 469--491.
\newblock \href {https://doi.org/10.1137/S0036142901389025}
  {\path{doi:10.1137/S0036142901389025}}.

\bibitem{franzelli2010}
B.~Franzelli, E.~Riber, M.~Sanjos\'e, T.~Poinsot, A two-step chemical scheme
  for kerosene-air premixed flames, Combustion and Flame 157~(7) (2010)
  1364--1373.
\newblock \href {https://doi.org/10.1016/j.combustflame.2010.03.014}
  {\path{doi:10.1016/j.combustflame.2010.03.014}}.

\bibitem{nambully2014}
S.~Nambully, P.~Domingo, V.~Moureau, L.~Vervisch, A filtered-laminar-flame
  {PDF} sub-grid scale closure for {LES} of premixed turbulent flames. {Part
  I}: Formalism and application to a bluff-body burner with differential
  diffusion, Combustion and Flame 161~(7) (2014) 1756--1774.
\newblock \href {https://doi.org/10.1016/j.combustflame.2014.01.005}
  {\path{doi:10.1016/j.combustflame.2014.01.005}}.

\bibitem{poinsot2005}
T.~Poinsot, D.~Veynante, Theoretical and Numerical Combustion, 2nd Edition,
  R.T. Edwards, Inc., 2005.

\end{thebibliography}
